# An Ordinal Bargaining Solution with Fixed-Point Property

**Dongmo Zhang**                                    DONGMO@SCM.UWS.EDU.AU
**Yan Zhang**                                        YAN@SCM.UWS.EDU.AU
*Intelligent Systems Laboratory*
*School of Computing and Mathematics*
*University of Western Sydney, Australia*

## Abstract

Shapley's impossibility result indicates that the two-person bargaining problem has no non-trivial ordinal solution with the traditional game-theoretic bargaining model. Although the result is no longer true for bargaining problems with more than two agents, none of the well known bargaining solutions are ordinal. Searching for meaningful ordinal solutions, especially for the bilateral bargaining problem, has been a challenging issue in bargaining theory for more than three decades. This paper proposes a logic-based ordinal solution to the bilateral bargaining problem. We argue that if a bargaining problem is modeled in terms of the logical relation of players' physical negotiation items, a meaningful bargaining solution can be constructed based on the ordinal structure of bargainers' preferences. We represent bargainers' demands in propositional logic and bargainers' preferences over their demands in total preorder. We show that the solution satisfies most desirable logical properties, such as individual rationality (logical version), consistency, collective rationality as well as a few typical game-theoretic properties, such as weak Pareto optimality and contraction invariance. In addition, if all players' demand sets are logically closed, the solution satisfies a fixed-point condition, which says that the outcome of a negotiation is the result of mutual belief revision. Finally, we define various decision problems in relation to our bargaining model and study their computational complexity.

## 1. Introduction

Bargaining has been a central research topic in economics for over five decades and has become an interesting issue in computer science in recent years (Osborne & Rubinstein, 1990; Rosenschein & Zlotkin, 1994; Muthoo, 1999). In his ground-breaking paper, Nash (1950) models a bargaining problem as a pair $(S, d)$, where $S \subseteq \Re^2$ is a subset of two-dimensional Euclidean space (feasible set), representing a set of utility pairs that can be derived from bargainers' preferences on feasible outcomes, and $d$ is a point in $D$ designated to be the "disagreement point". A bargaining solution is then a function that assigns each bargaining problem $(S, d)$ a point in $S$ (Thomson, 1994).

The Nash bargaining model presumes that bargainers' preferences are represented by von Neumann-Morgenstern utility, referred to as *cardinal utility* (Myerson, 1991). Under such an assumption, two utility functions can be viewed as the same if one can be derived from the other by an affine positive transformation. Thus a bargaining solution based on Nash's bargaining model should be invariant under any affine positive transformations. However, traditional economic theory considers bargaining problems in which players' preferences are represented in ordinal (Calvo & Peters, 2005). Therefore, ideally, a bargaining solution should be invariant under any order-preserving transformations on utilities. This property is referred to as *ordinal invariance* in the game-theoretic literature (Thomson, 1994). A bargaining solution that possesses such a property is called an *ordinal solution*. Obviously, ordinal bargaining solutions are more desirable than cardinal solutions because ordinal information about players' preferences is easier to elicit than *cardinal* preferences and the corresponding solutions can be more robust (Sakovics, 2004; Calvo & Peters, 2005). However, none of the well known bargaining solutions (Nash, 1950; Kalai & Smorodinsky, 1975; Kalai, 1977; Perles & Maschler, 1981) are ordinal. In fact, Shapley (1969) showed that for the two-person bargaining problem (bilateral bargaining) there is no non-trivial (i.e., strongly individual rational) ordinal solution[1]. The result is generally referred to as *Shapley's impossibility result* in the game-theoretic literature.

Shapley's negative result obviously discouraged the investigation of ordinal bargaining, notwithstanding Shapley himself demonstrated ten years later that ordinal solutions exist for the three-person bargaining problem (Shubik, 1982). The study of ordinal bargaining theory did not regain the focus of game theory until very recently. Kibris (2001) provided an axiomatic characterization for an ordinal solution of the three-person bargaining problem based on Nash's bargaining model[2]. Safra and Samet (2004) extended the result to the bargaining problems with more than three players. Rubinstein et al. (1992) and O'Neill et al. (2004) investigated the ordinal bargaining problem by varying Nash's bargaining model. Calvo and Perers (2005) explored the problem of ordinal bargaining in which at least one player is cardinal. Nevertheless, the problem of ordinal bargaining is still considered to be an unsolved problem. Most of these pieces of work focus on the existence of ordinal solutions. None of these proposed solutions gains strong intuitive support. Looking for meaningful ordinal bargaining solutions is still an outstanding problem in game theory (Sakovics, 2004). To show the difficulty of ordinal bargaining, let us consider a simple but typical bargaining scenario:

**Example 1** (Muthoo, 1999) *Two players, A and B, bargain over the partition of a cake. Let $x_i$ be the share of the cake in percentage to player $i$ ($i = A, B$). The set of possible agreements is represented by $\Omega = \{(x_A, x_B) : 0 \leq x_A \leq 100 \text{ and } x_B = 100 - x_A\}$. For each $x_i \in [0, 100]$, $u_i(x_i)$ is player $i$'s utility from obtaining a share $x_i$ of the cake.*

Assume that player A has a linear utility scale of its share, $u_A(x_A) = x_A$, and player B has a utility scale that is proportional to the square of his share, $u_B(x_B) = x_B^2$. Failure to agree is rated 0 by both A and B. Consider two most influential bargaining solutions: *Nash's solution* (Nash, 1950) and *Kalai-Smorodinsky's solution* (Kalai & Smorodinsky, 1975). It is easy to calculate that Nash's bargaining solution to the problem gives the outcome (33.3, 66.7) and Kalai-Smorodinsky's solution gives (38.2, 61.8). Both solutions are in favor of player B. This is because player B is less risk-averse (has concave utility) than player A (has linear utility). For both Nash's solution and Kalai-Smorodinsky's solution, risk-loving players has advantage in bargaining comparing to risk-neutral and risk-averse players (see Roth's book, 1979a, p.35-60). Now consider an order-preserving transformation $\tau(x) = \sqrt{x}$ on player B's utility. The transformed utility of player B becomes linear. Under the new utility scales, both Nash's solution and Kalai-Smorodinsky's solution give $(50, 50)$ as outcome. This means that none of the solutions is ordinally invariant.

---

1. See the work of Thomson (1994) for an easy proof.
2. The original work was not formally published. Kibris (2004) gave a brief note.

This example clearly shows that the non-linearity of utility functions, which expresses the risk posture of a player, determines the outcomes of bargaining but collapses under ordinal transformations. In other words, ordinal transformations filter out useful information that is expressible by cardinal utility but not expressible by ordinal utility. This explains why "*no resolution of the two-player bargaining problem can be made on the basis of ordinal utility alone .... A satisfactory theory of bilateral bargaining requires knowledge of something more than just an ordering of the bargainers' preferences*" (see Shubik's book, 1982, p.94-98).

All in all, ordinal preference is insufficient to fully specify a bargaining situation. A bargaining model must supply a way to express the information additional to ordinal preferences, such as bargainers' attitude towards risk. This article aims to demonstrate with an ordinal solution to the bilateral bargaining problem that the language of logic can be used to express the knowledge that is required for modeling bargaining with ordinal preferences.

In recent years, the studies on logic-based frameworks of bargaining and negotiation have received considerable attention in the field of artificial intelligence (AI) (Sycara, 1990; Kraus, Sycara, & Evenchik, 1998; Parsons, Sierra, & Jennings, 1998; Zhang, Foo, Meyer, & Kwok, 2004; Meyer, Foo, Kwok, & Zhang, 2004; Zhang, 2005, 2007). These frameworks utilize logical languages to represent bargaining situations so that physical negotiation items, bargaining conflicts, players' beliefs and mutual threats can be explicitly expressed, which differentiates themselves from the traditional game-theoretic models. In this paper we propose an ordinal solution to the bilateral bargaining problem based on the logical model introduced by Zhang and Zhang (2006a).

The organization of the paper is the following. Section 2 presents the formal model of bargaining. We will use a finite propositional language to describe bargainers' demands. The bargainers' preferences on their demands are sorted in total preorder. A bargaining problem is then defined as a pair of hierarchies of two parties' demand sets. The construction of the bargaining solution, presented in Section 3, is based on the idea that each party tries to maximize their prior demands to be included in the final agreement while keeping the outcome to be consistent. The approach can be viewed as an extension of Nebel's prioritized base revision to the two-agent setting (Nebel, 1992). Section 4 and Section 5 are devoted to the discussions on the properties of the proposed solution. We shall prove in Section 4 that the solution satisfies most desirable logical properties, such as the logical version of Individual Rationality, Consistency and Collective Rationality. More extraordinarily, we shall show that the solution satisfies a desirable fixed-point condition introduced by Zhang et al. (2004), which says that the outcome of bargaining is the result of mutual belief revision. Section 5 focuses on the discussion of the game-theoretic properties of the solution. We prove that the solution satisfies Weak Pareto Optimality, Restricted Symmetry and Contraction Invariance. Section 6 is devoted to a discussion of how bargainers' attitudes towards risk are represented in our model and how they determine players' bargaining power. Section 7 investigates the complexity issues related to the proposed model. We consider four major decision problems in relation to a bargaining game and provide their computational complexity results. The final two sections conclude the work with a discussion of the related work.

The paper is made to be self-contained. However, the reader will find that basic knowledge in belief revision and game-theoretic bargaining theory can be helpful for a better understanding of the concepts introduced in the paper. For an introductory survey of the areas, see Gärdenfors's article (1992) and Thomson's article (1994), respectively.

## 2. Representation of Bargaining Problems

As we have seen in Example 1, an ordinal solution to the bilateral bargaining problem requires information in addition to ordinal preferences. In order to express such extra information, we model a bargaining situation in two aspects: the *physical bargaining terms*, described in propositional logic, and the *ordering of the bargainers' preferences* on their bargaining terms, described in total preorder. Since the bilateral bargaining is the most challenging problem to bargaining theory, we shall restrict ourselves to the bargaining problem with only two players.

### 2.1 Preliminaries

We assume that each party has a set of negotiation items, referred to as *demand set*, described by a finite propositional language $\mathcal{L}$. The language consists of a finite set of propositional variables and the propositional connectives $\neg$, $\vee, \wedge$, $\rightarrow$ and $\leftrightarrow$ with standard syntax and semantics. The logical closure operator $Cn$ is defined as $Cn(X) = \{\varphi : X \vdash \varphi\}$, where $X$ is a set of sentences. We say $X$ to be *logically closed* or a *belief set* if $X = Cn(X)$. If $X$ and $Y$ are two sets of sentences, $X + Y$ denotes $Cn(X \cup Y)$.

Suppose that $X_1$ and $X_2$ are two sets of sentences. To simplify exploration, we use $X_{-i}$ to represent the other set among $X_1$ and $X_2$ if $X_i$ is one of them. If $D$ is a vector of two components, $D_1$ and $D_2$ will represent each component of $D$.

### 2.2 Bargaining Games

As shown in the previous section, cardinal utility encodes two types of information: *bargainers' attitude towards risk* (via non-linearity of utility functions) and *bargainers' preferences on possible outcomes* (via the ordering of utility values). One may ask whether a theory of bargaining can be based purely on ordinal information about preferences. In order to investigate such a possibility, Osborne and Rubinstein introduced a different way to represent a bargaining situation, with which the preference information is separated from the physical bargaining terms (Osborne & Rubinstein, 1990). Precisely, they define a bargaining problem as a four-tuple $(X, D, \geq_1, \geq_2)$, where $X$ is a set of feasible outcomes (in physical terms), $D$ is the disagreement event, and $\geq_i$ is a complete transitive reflexive ordering over the set $X \cup D$, representing bargainer $i$'s preferences. As shown by Example 1, however, simply describing physical bargaining terms without specifying their relations does not suffice to lead to an ordinal solution (see also Osborne & Rubinstein's book, 1990, p.32). In this paper, we further extend Osborne & Rubinstein's model in such a way that the physical negotiation items are represented in logical formulae.

**Definition 1** *A bargaining game is a pair $((X_1, \succeq_1), (X_2, \succeq_2))$, where $X_i$ ($i = 1, 2$) is a logically consistent set of sentences in $\mathcal{L}$ and $\succeq_i$ is a complete transitive reflexive order (total preorder or weak order) over $X_i$ that satisfies the following logical constraint[3]:*

**(LC)** *If $\varphi_1, \cdots, \varphi_n \vdash \psi$, then there is $k$ ($1 \leq k \leq n$) such that $\psi \succeq_i \varphi_k$.*

*We call the pair $(X_i, \succeq_i)$ the prioritized demand set of player $i$. For any $\varphi, \psi \in X_i$, $\psi \succ_i \varphi$ denotes that $\psi \succeq_i \varphi$ and $\varphi \nsucceq_i \psi$. $\psi \approx_i \varphi$ denotes that $\psi \succeq_i \varphi$ and $\varphi \succeq_i \psi$.*

Intuitively, a bargaining game is the formal representation of a bargaining situation in which each player describes his demands in logical formulae and expresses his preferences on his demands in total preorder. We assume that each player has consistent demands. The preference ordering of each player reflects the *degree of entrenchment* in which the player defends his demands. The logical constraint LC says that if the demand $\psi$ is a logical consequence of the demands $\varphi_1, \cdots, \varphi_n$, then $\psi$ should not be less entrenched than all the $\varphi_i$ because if you fail to defend $\psi$, at least one of the $\varphi_i$ has to be dropped (otherwise you would not have lost $\psi$). It is easy to see that such an ordering is similar to Gädenfors and Makinson's epistemic entrenchment (Gärdenfors & Makinson, 1988). In fact, the logical constraint LC, introduced by Zhang and Foo (2001), is actually the combination of the postulates EE2 and EE3 of epistemic entrenchment ordering.

**Observation 1** *Let $\succeq$ be a total preorder on $X$. The logical constraint LC is equivalent to the conjunction of the following conditions:*

1. *If $\varphi \vdash \psi$, then $\psi \succeq \varphi$.*

2. *Either $\varphi \wedge \psi \succeq \varphi$ or $\varphi \wedge \psi \succeq \psi$.*

**Proof:** *It is easy to verify that LC implies these two conditions. We now prove that the conditions imply LC by induction on $n$. Obviously, LC holds when $n = 1$. Assume that LC holds when $n = l$. Suppose that $\varphi_1, \cdots, \varphi_{l-1}, \varphi_l, \varphi_{l+1} \vdash \psi$, then $\varphi_1, \cdots, \varphi_{l-1}, (\varphi_l \wedge \varphi_{l+1}) \vdash \psi$. By the inductive assumption, either $\psi \succeq \varphi_l \wedge \varphi_{l+1}$ or there is $k$ ($1 \leq k \leq l-1$) such that $\psi \succeq \varphi_k$. If $\psi \succeq \varphi_l \wedge \varphi_{l+1}$, by the condition 2 and transitivity of $\succeq$, either $\psi \succeq \varphi_l$ or $\psi \succeq \varphi_{l+1}$. Therefore in any case there is $k$ ($1 \leq k \leq l+1$) such that $\psi \succeq \varphi_k$.* ¶

We remark that the preference ordering represents how firmly an agent entrenches her demands rather than her gain or payoff[4]. For instance, suppose that $p_1$ represents the demand of a seller "*the price of the good is no less than \$10*" and $p_2$ denotes "*the price of the good is no less than \$8*". Obviously the seller could get higher payoff from $p_1$ than $p_2$. However, since $p_1$ implies $p_2$, she will entrench $p_2$ no less firmly than $p_1$, *i.e.*, $p_2 \succeq p_1$,

---

3. A complete transitive reflexive order, i.e., total preorder or weak order, satisfies the following properties:

- Completeness or totality: $\varphi \succeq \psi$ or $\psi \succeq \varphi$.
- Reflexivity: $\varphi \succeq \varphi$.
- Transitivity: if $\varphi \succeq \psi$ and $\psi \succeq \chi$ then $\varphi \succeq \chi$.

4. We will define bargainers' gains in Section 5.1.

because, if she fails to keep $p_1$, she can still bargain for $p_2$ but the loss of $p_2$ means the loss of both.

There is another significant difference between our model of bargaining and the game-theoretic model. The game-theoretic model abstracts a bargaining situation into a numerical game. The demands, preferences and risk posture of a player are all represented in utility values. In our model, however, these factors are abstracted into logical statements and ordering on these statement, i.e., a prioritized demand set. Note that we have endowed the word "demand" with a broad meaning. In our model, a demand of a player can be everything that is related to the negotiation and that the player wants to keep in the final agreement. It can be a physical item the player wants to obtain from the other party. It can also be a piece of knowledge, a belief, a goal, a constraint or even a thread, whatever, the player wants to keep in the agreement. Different from the logics for rational agency, such as the BDI logics, we do not distinguish knowledge, belief, goal from the "real" demands. For instance, a logical tautology can be a demand of a player if the player considers that it should be included in the final agreement. Consider Example 1 again. Typical demands of player A are $x_A \geq 40$, $x_A \geq 50$, $x_A \geq 60$, and so on[5], which mean that the player wants to get a share of the cake no less than 40%, 50%, 60%, $\cdots$. Player B's demands are the opposite, say $x_B \geq 40$, $x_B \geq 50$, $x_B \geq 60$, $\cdots$. Besides these "real" demands, there are a set of domain constraints in the players' demand sets, such as $x_A + x_B \leq 100$, $x_A \geq y \rightarrow x_A \geq z$ and $x_B \geq y \rightarrow x_B \geq z$ whenever $y \geq z$. These constraints link all the demands of the players together, therefore, play an important rule in the determination of bargaining outcomes. For instance, we cannot have both $x_A \geq 55$ and $x_B \geq 55$ in the final agreement because they are inconsistent with the constraints. One may wonder what if a player does not include these constraints in her demand set. In this case, the player would have to accept unreasonable results, such as $x_A \geq 55$ and $x_B \geq 55$. In other words, if "*we idealize the bargaining problem by assuming that the two individuals are highly rational*" (see the work of Nash, 1950, p.155), we should assume that these constraints and background knowledge are included in the demand set of each player (see more discussions on this example in Section 6). Again, the ordering of demand sets does not reflect the gains of a player as we mentioned above. It is very likely that a rational agent could give her highest priority to the above mentioned constraints and background knowledge since she might be never going to give up these fundamental rules.

The following example shows that our model is more suitable for discrete domain of problems.

**Example 2** *A couple are making their family budget for the next year. The husband wants to change his car to a new fancy model and have a domestic holiday. The wife is going to implement her dream of a romantic trip to Europe and suggests to redecorate their kitchen. Both of them know that they can't have two holidays in one year. They also realize that they cannot afford a new car and an overseas holiday in the same year without getting a loan from the bank. However, the wife does not like the idea of borrowing money.*

In order to represent the situation in logic, let $c$ denote "buy a new car", $d$ stand for "domestic holiday", $o$ for "overseas holiday", $k$ for "kitchen redecoration" and $l$ for "loan".

---

5. Note that the player may have different demands at different stages of negotiation.

Then $\neg(d \wedge o)$ means that it is impossible to have both domestic holiday and overseas holiday. The statement $(c \wedge o) \rightarrow l$ says that if they want to buy a new car and also have an overseas holiday, they have to get a loan from the bank.

With the above symbolization, we can express the husband's demands in the following set:

$$X_1 = \{c, d, \neg(d \wedge o), (c \wedge o) \rightarrow l\}$$

Similarly, the wife's demands can be represented by:

$$X_2 = \{o, k, \neg(d \wedge o), (c \wedge o) \rightarrow l, \neg l\}$$

Let us assume that the husband's preferences over his demands are:

$$\neg(d \wedge o) \approx_1 (c \wedge o) \rightarrow l \succ_1 c \succ_1 d$$

and the wife's preferences are:

$$\neg(d \wedge o) \approx_2 (c \wedge o) \rightarrow l \succ_2 o \succ_2 k \succ_2 \neg l$$

Note that both agents give the common beliefs: $\neg(d \wedge o)$ and $(c \wedge o) \rightarrow l$ the highest priority. This reflects that the couple are rational. We shall see that these common beliefs play important role in the determination of bargaining solution.

From the above example, we can also see the differences between our model and Osborne & Rubinstein's model (Osborne & Rubinstein, 1990). First, the demand sets $X_1$ and $X_2$ in our model represent players' physical demands rather than feasible outcomes (note that the two players share the same feasible set $X$ in Osborne & Rubinstein's model). In our model, the feasible set is generated from the demand sets through a procedure of conflict resolving (see Definition 2 and Example 3). Secondly, we do not have a representation of disagreements. If a negotiation ends with disagreement, we simply assume that the agreement is empty. Thirdly, in Osborne & Rubinstein's model, the items in $X$ are independent, while in our model, the items from both players' demand sets are related through their logical relations and the orderings (via LC). This allows us to represent bargainers' attitude towards risk.

## 2.3 Hierarchy of Demands

Before we construct our bargaining solution, we present the basic properties of prioritized demand sets. Consider a prioritized demand set $(X, \succeq)$ for a single agent. We define recursively a hierarchy, $\{X^j\}_{j=1}^{+\infty}$, of $X$ with respect to the ordering $\succeq$ as follows:

1. $X^1 = \{\varphi \in X : \forall \psi \in X(\varphi \succeq \psi)\}$; $T^1 = X \backslash X^1$.

2. $X^{j+1} = \{\varphi \in T^j : \forall \psi \in T^j(\varphi \succeq \psi)\}$; $T^{j+1} = T^j \backslash X^{j+1}$.

The intuition behind the construction is the following: at each stage of the construction, we collects all the maximal elements from the current demand set and remove them from the set for the next stage of the construction. It is easy to see that there exists a number $n$ such that $X = \bigcup\limits_{j=1}^{n} X^j$ due to the logical constraint LC[6]. Therefore a demand set $X$ can be always written as $X^1 \cup \cdots \cup X^n$, where $X^j \cap X^k = \emptyset$ for any $j \neq k$ (see Figure 1). Also for any $\varphi \in X^j$ and $\psi \in X^k$, $\varphi \succ \psi$ if and only if $j < k$. In other words, the prioritized demand set $(X, \succeq)$ uniquely determines a partition of $X$, $\bigcup\limits_{j=1}^{n} X^j$, with a total order over the partition. Therefore, the concept of prioritized demand set is equivalent to the concept of so-called nicely-ordered partition of a belief set[7], introduced by Zhang and Foo (Zhang & Foo, 2001), if the demand set is logically closed.

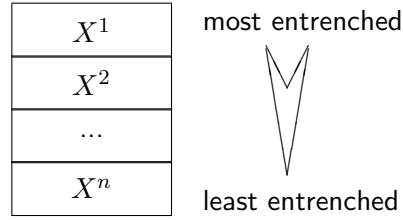

Figure 1: *The hierarchy of a demand set.*

In the sequent, we write $X^{\leq k}$ to denote $\bigcup\limits_{j=1}^{k} X^j$.

## 2.4 Prioritized Base Revision

Once we have a hierarchy of the demand set for each agent, we are able to define a belief revision operator for each agent by following Nebel's *prioritized base revision* (Nebel, 1992)[8]. We define a revision function $\otimes$ as follows:

For any demand set $(X, \preceq)$ and a set $F$ of sentences,

$$X \otimes F \stackrel{def}{=} \bigcap_{H \in X \Downarrow F} (H + F),$$

where $X \Downarrow F$ is defined as: $H \in X \Downarrow F$ if and only if

1. $H \subseteq X$,

2. for all $k$ $(k = 1, 2, \cdots)$, $H \cap X^k$ is a maximal subset of $X^k$ such that $\bigcup\limits_{j=1}^{k} (H \cap X^j) \cup F$ is consistent.

In other words, $H$ is a maximal subset of $X$ that is consistent with $F$ and gives priority to the higher ranked items. The following result will be used in Section 4.

**Lemma 1** (Nebel, 1992) *If $X$ is logically closed, then $\otimes$ satisfies all AGM postulates.*

---

6. Note that $X$ can be an infinite set even though the language is finite.

7. A nicely-ordered partition of a belief set $K$ is a triple $(K, \wp, \geq)$, where $\wp$ is a partition of $K$ and $\geq$ is a total order on $\wp$ that satisfies the logical constraint LC. See the work of Zhang and Foo (2001) p.540.

8. The idea of the construction can be traced back to Poole and Brewka's approach to default logic (Brewka, 1989; Poole, 1988).

## 3. Bargaining Solution

In Nash's bargaining model, a bargaining solution is defined as a function that assigns to each bargaining game a point in the feasible set of the game. However, we do not have such a simple definition because in our model the set of possible agreements (feasible set) is not given in a bargaining game. We have to generate the feasible set from the demand sets. This involves a process of conflict resolving.

### 3.1 Possible Deals

Whenever the demands from two agents conflict, at least one agent has to make a concession in order to reach an agreement. The simple way of making a concession is to withdraw a number of demands. In such sense, a possible agreement is a pair of subsets of two players' original demand sets such that the collection of remaining demands is consistent. Obviously each player would like to keep as many original demands as possible. In addition, if a player has to give up a demand, the player typically gives up the one with the lowest priority. This idea leads to the following definition of possible deals.

**Definition 2** *Let $G = ((X_1, \succeq_1), (X_2, \succeq_2))$ be a bargaining game. A deal of $G$ is a pair $(D_1, D_2)$ satisfying the following conditions: for each $i = 1, 2$,*

1. *$D_i \subseteq X_i$;*

2. *$X_1 \cap X_2 \subseteq D_i$;*

3. *for each $k$ $(k = 1, 2, \cdots)$, $D_i \cap X_i^k$ is a maximal subset of $X_i^k$ such that $\bigcup_{j=1}^{k} (D_i \cap X_i^j) \cup D_{-i}$ is consistent.*

*where $\{X_i^j\}_{j=1}^{+\infty}$ is the hierarchy of $X_i$ defined in Section 2.3. The set of all deals of $G$ is denoted by $\Omega(G)$, called the feasible set of the game.*

This definition is obviously an analogue of Nebel's notion $\Downarrow$ (see Section 2.4). The only difference is that the procedure of maximization here is interactive between two agents: *given one player's demands, the other player always tops up his demands with the highest prioritized items provided the overall outcome is consistent.*

Since we have assumed that both $X_1$ and $X_2$ are consistent, $\Omega(G)$ is non-empty. Specifically, if $X_1 \cup X_2$ is consistent, we have $\Omega(G) = \{(X_1, X_2)\}$.

**Example 3** *Consider the bargaining game in Example 2. According to the preference orderings of the couple, the game has three possible deals:*
*$D^1 = (\{\neg(d \wedge o), (c \wedge o) \to l, c, d\}, \{\neg(d \wedge o), (c \wedge o) \to l, k, \neg l\})$.*
*$D^2 = (\{\neg(d \wedge o), (c \wedge o) \to l, c\}, \{\neg(d \wedge o), (c \wedge o) \to l, o, k\})$.*
*$D^3 = (\{\neg(d \wedge o), (c \wedge o) \to l\}, \{\neg(d \wedge o), (c \wedge o) \to l, o, k, \neg l\})$.*
*Therefore $\Omega(G) = \{D^1, D^2, D^3\}$.*

## 3.2 The Core of Agreement

We have shown how to generate possible deals, which form the set of possible agreements, i.e., the feasible set, by resolving conflicting demands. Now we are at the same level as the game-theoretic model that, to define a bargaining solution, we only have to select a deal from all possible deals. Obviously, if the demands from two parties contradict, there are multiple possible deals. Different deals would be in favor of different parties. For instance, in Figure 2, the deal $D'$ is in favor of player 1 while $D''$ is in favor of player 2. Therefore, the conflicts in choosing outcomes still exist. The major concern of a bargaining theory is how to measure and balance the gain of each negotiating party.

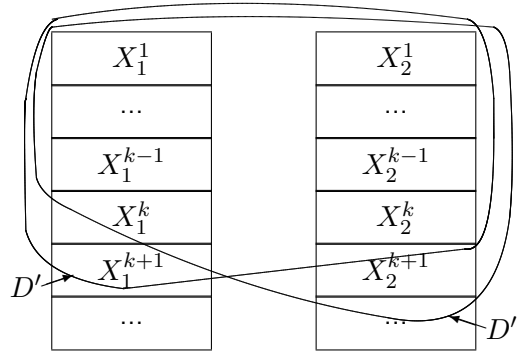

Figure 2: *Different deals are in favor of different players.*

Instead of counting the number of demands that each party can remain from a deal, we consider the top block demands that a player keeps in the deal (the top levels of demands in each player's demand hierarchy) and ignore all the demands that are not included in the top blocks for the purpose of measuring players' gains.

Given a deal $D$, we shall use the maximal number of the top levels of demands in the deal as the indicator of each player's gain from the deal, i.e., $\max\{k : X_i^{\leq k} \subseteq D_i\}$ for $i = 1, 2$. For instance, in Figure 2, player 1 remains maximally the top $k - 1$ levels of her demands from the deal $D''$ while player 2 can successfully gain the top $k + 1$ levels of his demands from the same deal. With deal $D'$, both players can remain the top $k$ levels of demands.

In order to compare different deals, we refer the gain index of a deal to the gain of the player whoever receives less from the deal, i.e., $\min\{\max\{k : X_1^{\leq k} \subseteq D_1\}, \max\{k : X_2^{\leq k} \subseteq D_2\}\}$, or equivalently, $\max\{k : X_1^{\leq k} \subseteq D_1 \text{ and } X_2^{\leq k} \subseteq D_2\}$. For instance, in Figure 2, the gain index of $D'$ is $k$ while the gain index of $D''$ is $k - 1$. Therefore we can say that $D'$ is better than $D''$ because, with $D'$, both players can remind at least top $k$ blocks in their demand hierarchies but $D''$ can't.

Formally, let

$$\pi_{max}^G = \max_{(D_1, D_2) \in \Omega(G)} \{k : X_1^{\leq k} \subseteq D_1 \text{ and } X_2^{\leq k} \subseteq D_2\} \tag{1}$$

and

$$\gamma(G) = \{(D_1, D_2) \in \Omega(G) : X_1^{\leq \pi_{max}^G} \subseteq D_1 \And X_2^{\leq \pi_{max}^G} \subseteq D_2\} \tag{2}$$

Then $\gamma(G)$ collects all the "best deals" from all the possible deals of $G$. This is because none of other deals can contain more than the top $\pi^G_{max}$ levels of demands from both players. Note that $\pi^G_{max}$ may be infinite when $X_1 \cup X_2$ is consistent. Let

$$(\Phi_1, \Phi_2) = (X_1^{\leq \pi^G_{max}}, X_2^{\leq \pi^G_{max}}) \tag{3}$$

We call $\Phi = (\Phi_1, \Phi_2)$ the *core of the game*. Therefore the core contains the top block demands that all the best deals contain, therefore should be included in the final agreement. The following lemma gives another way to calculate the core of a game.

**Lemma 2** $\pi^G_{max} = \max\{k : X_1^{\leq k} \cup X_2^{\leq k} \cup (X_1 \cap X_2)$ *is consistent*$\}$.

**Proof:** Let $\pi = \max\{k : X_1^{\leq k} \cup X_2^{\leq k} \cup (X_1 \cap X_2)$ is consistent$\}$. It is easy to show that $X_1^{\leq \pi^G_{max}} \cup X_2^{\leq \pi^G_{max}} \cup (X_1 \cap X_2)$ is consistent because $\gamma(G)$ is non-empty. Therefore $\pi^G_{max} \leq \pi$. On the other hand, since $X_1^{\leq \pi} \cup X_2^{\leq \pi} \cup (X_1 \cap X_2)$ is consistent, there exists a deal $(D_1, D_2) \in \Omega(G)$ such that $X_i^{\leq \pi} \subseteq D_i$ and $X_1 \cap X_2 \subseteq D_i$ for each $i = 1, 2$. Thus $\pi \leq \pi^G_{max}$. We conclude that $\pi = \pi^G_{max}$. ¶

### 3.3 The Solution

At first sight, a bargaining solution can be easily defined as a function that assigns each bargaining game $G$ a possible deal in $\Omega(G)$. More likely, the solution could select one of the "best deals" from $\gamma(G)$. However, due to the multiplicity of $\gamma(G)$ (see Figure 3), such a selection is not always feasible if we want the solution to be symmetric to each player. To show the difficulty, let us consider the following example.

**Example 4** *Consider a bargaining game $G = ((X_1, \succeq_1), (X_2, \succeq_2))$, described in three propositional variables $p, q$ and $r$, where $X_1 = \{\{p\}, \{r\}\}$ and $X_2 = \{\{q\}, \{\neg r\}\}$ (note that the demand sets are represented in the form of hierarchy where $p \succ_1 r$ and $q \succ_2 \neg r$ (see Section 2.3)). It is easy to know that the game has two possible deals, $(\{p, r\}, \{q\})$ and $(\{p\}, \{q, \neg r\})$, which are all in $\gamma(G)$. However, none of the deals can lead to an impartial solution. A reasonable solution to the problem should be $(\{p\}, \{q\})$, which take the intersection of all the best deals for each player, respectively.*

Base on the above intuitive explanation, we are now ready to present our bargaining solution.

**Definition 3** *The bargaining solution is the function $F$ defined as follows, which maps a bargaining game $G = ((X_1, \succeq_1), (X_2, \succeq_2))$ to a pair of sets of sentences:*

$$F(G) \stackrel{def}{=} (\bigcap_{(D_1, D_2) \in \gamma(G)} D_1, \bigcap_{(D_1, D_2) \in \gamma(G)} D_2) \tag{4}$$

*where $\gamma(G)$ is defined by Equation (2).*

For a better understanding of the construction of our solution, we would like to make the following remarks:

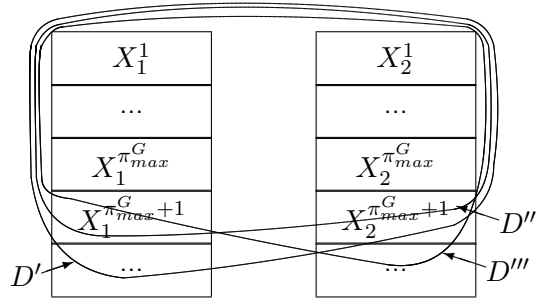

Figure 3: *Multiplicity of best deals.*

1. The solution gives a prediction of bargaining outcome for each game. Given a game $G$, $F_i(G)$ represents all the demands the player $i$ can successfully remain at the end of bargaining. It also means that these demands are accepted by the other player. Therefore the final agreement of the bargaining can be defined as $F_1(G) \cup F_2(G)$[9].

2. Note that $F_i(G) \subseteq X_i$ for each $i = 1, 2$. Therefore the solution means a compromise to each player. Both players may make concessions to their demands in order to reach the agreement. Obviously, if there is no conflict between the players' original demands, i.e., $X_1 \cup X_2$ is consistent, no concession is needed, that is, $F_i(G) = X_i$ ($i = 1, 2$).

3. The construction of the solution takes a skeptical view in the sense that, for each player, a demand item is included in the solution only if it belongs to all the "best deals". In other words, the solution gives only a cautious prediction of bargaining outcome. As a result, the solution is not necessarily a deal if there are multiple elements in $\gamma(G)$.

4. The solution is unique to each bargaining game. Like the bargaining problem with non-convex domain, we have to scarify strict Pareto optimality to gain the uniqueness (this is the reason we take cautious prediction). However, we will show that the solution is weakly Pareto optimal (see Section 5.2 for more discussions).

5. In our bargaining model, we do not specify disagreement points. In fact, we assume that if the solution gives an empty agreement, i.e., $F(G) = (\emptyset, \emptyset)$, then the negotiation reaches a disagreement. In other words, $(\emptyset, \emptyset)$ is the default disagreement point for any bargaining game.

**Example 5** *Continue on Example 3. According to the hierarchies of the demand sets shown in Example 2, the core of the game is:*

$$(\{\neg(d \wedge o), (c \wedge o) \to l, c\}, \{\neg(d \wedge o), (c \wedge o) \to l, o\})$$

---

9. Alternatively, we can define the final agreement as $Cn(F_1(G) \cup F_2(G))$ if we consider that the outcome of the negotiation contains all the logical consequences of the demands in the agreement. In addition, the relation of the items in the agreement should be read as "and" rather than "or". In other words, all items in the agreement are accepted by all players.

*Therefore $\gamma(G)$ contains only a single deal, which is $D^2$ (see Example 3). The solution is then*

$$F(G) = D^2 = (\{\neg(d \wedge o), (c \wedge o) \to l, c\}, \{\neg(d \wedge o), (c \wedge o) \to l, o, k\})$$

*In words, the couple agree upon the commonsense that they can only have one holiday and they have to get a loan if they want to buy a new car and to go overseas for holiday. The husband accepts his wife's suggestion to have holiday in Europe and the wife agrees on buying a new car.*

*Now consider the following preference orderings:*

$$\neg(d \wedge o) \approx_1 (c \wedge o) \to l \succ_1 d \succ_1 c$$

$$\neg(d \wedge o) \approx_2 (c \wedge o) \to l \succ_2 o \succ_2 k \succ_2 \neg l$$

*Therefore the demand hierarchies become:*
*$X_1 = \{\{\neg(d \wedge o), (c \wedge o) \to l\}, \{d\}, \{c\}\}$.*
*$X_2 = \{\{\neg(d \wedge o), (c \wedge o) \to l\}, \{o\}, \{k\}, \{\neg l\}\}$.*
*Let $G'$ denote the game. The deals are the same as the original hierarchy as shown in Example 3. However the solution of the game becomes $(\{\neg(d \wedge o), (c \wedge o) \to l\}, \{\neg(d \wedge o), (c \wedge o) \to l, k\})$ because $\pi_{max}^{G'} = 2$ and all three deals are included in $\gamma(G')$. Note that the final agreement does not include the demands which lead to conflicting (d, o, ¬l) but keeps the demands which do not lead to conflicting (k). In other words, the solution excludes the demands that lead to a conflict but keeps the demands that are not involved in any conflicts even though they are in low priorities.*

## 4. Logical Properties of the Bargaining Solution

In the following two sections, we discuss the properties of the bargaining solution introduced in the previous section. According to Zhang (2007), a bargaining solution satisfies the axioms *Collective Rationality*, *Scale Invariance*, *Symmetry* and *Mutually Comparable Monotonicity* as well as the basic assumptions *Individual Rationality*, *Consistency* and *Comprehensiveness* if and only if it is the logical version of Kalai-Smorodinsky solution (Kalai & Smorodinsky, 1975). Among these properties, *Collective Rationality*, *Individual Rationality* and *Consistency* capture the logical properties of a bargaining solution. *Scale Invariance*, *Symmetry* and *Mutually Comparable Monotonicity* reflect the game-theoretic properties of a bargaining solution. *Comprehensiveness* is an idealized assumption for a logic-based bargaining solution. Although there is a significant difference between the bargaining solution we defined in this paper and the one in that work, these two solutions share most desirable properties of bargaining solutions.

### 4.1 Generic Properties of Logic-Based Bargaining Solutions

It is easy to see that the solution we constructed in the previous section satisfies the following generic properties of a logic-based bargaining solution.

**Theorem 1** *For any bargaining game $G = ((X_1, \succeq_1), (X_2, \succeq_2))$, let $F(G) = (F_1(G), F_2(G))$. Then*

1. $F_1(G) \subseteq X_1$ and $F_2(G) \subseteq X_2$.                    *(Individual Rationality)*

2. $F_1(G) \cup F_2(G)$ is consistent.                    *(Consistency)*

3. If $X_1 \cup X_2$ is consistent, $F_i(G) = X_i$ for all $i$.                    *(Collective Rationality)*

**Proof:** The proofs for these properties are straightforward from the definition of the bargaining solution (Definition 3).                    ¶

Note that the logical version of Individual Rationality (IR) has a different meaning of its game-theoretic version. The logical version of IR means that each player concerns only her own demands, i.e., whether and how many of her demands are included in the final agreement. In contrast, the game-theoretic version of IR concerns about whether each player can gain no less than disagreement point from a negotiation (see more details in Section 5.1). The other two properties are quite intuitive.

The following example shows that our solution does not satisfy comprehensiveness, which requires that $\varphi \in F_i(G)$ and $\psi \succeq_i \varphi$ implies $\psi \in F_i(G)$ for each $i$ (see the work of Zhang, 2007).

**Example 6** *Consider a bargaining situation in which player 1's demand set is $X_1 = \{p, q\}$ and player 2's demand set is $X_2 = \{\neg p, r\}$, where $p, r, q$ are propositional variables. Assume that each player ranks her demands in same level (i.e., both demand sets have a singleton partition). Based on the assumption, it is easy to know that the solution of the game is $(\{q\}, \{r\})$. Therefore, the solution is not comprehensive (for instance, $q \preceq p$ and $q \in F_1(G)$ but $p \notin F_1(G)$).*

Since the solution does not satisfy comprehensiveness, according to Zhang (2007), it is not the logical version of Kalai-Smorodinsky solution. However, this does not mean that our solution is less intuitive. Although comprehensiveness is a common restriction in belief revision and game theory, it is by no means a desirable property of bargaining solution. In the above example, $q$ and $r$ are not involved in the conflict of the underlying bargaining game. Thus it is reasonable for the players to keep these irrelevant demands. In addition, the solution is syntax-dependent. If we represent the demand set as $X_1 = \{p \wedge q\}$ and $X_2 = \{\neg p \wedge r\}$, the solution will be $(\emptyset, \emptyset)$.

## 4.2 Fixed-Point Property

Besides the generic properties, the solution possesses another extraordinary logical property: the fixed-point property. We consider bargaining or negotiation as mutual persuasion: one persuades the other to accept her demands. The outcome of negotiation is then the result of mutual belief revision (Zhang et al., 2004). If it is the case, the negotiation outcome should satisfy the following fixed-point property.

**Theorem 2** *For any bargaining game $G = ((X_1, \succeq_1), (X_2, \succeq_2))$, if $X_1$ and $X_2$ are logically closed, the bargaining solution $F(G)$ satisfies the following fixed-point condition:*

$$F_1(G) + F_2(G) = (X_1 \otimes_1 F_2(G)) \cap (X_2 \otimes_2 F_1(G)) \tag{5}$$

*where $\otimes_i$ is the prioritized revision operator for player $i$ (see the definition in Section 2.4).*

Assume that $X_1$ and $X_2$ are two belief sets (so logically closed), representing the belief states of two agents. Mutual belief revision between the agents means that each agent takes part of the other agent's beliefs to revise his belief set. For instance, if $\Psi_1$ is a subset of $X_1$ and $\Psi_2$ is a subset of $X_2$, then $X_1 \otimes_1 \Psi_2$ is the revised belief set of player 1 after he accepts player 2's beliefs $\Psi_2$ while $X_2 \otimes_2 \Psi_1$ is the resulting belief set of player 2 after accepting $\Psi_1$. Such an interaction of belief revision can continue until it reaches a fixed point where the beliefs in common, i.e., $(X_1 \otimes_1 \Psi_2) \cap (X_2 \otimes_2 \Psi_1)$, are exactly the beliefs that the agents mutually accept, that is, $\Psi_1 + \Psi_2$. Note that each agent uses his own way of revision to rebuilt his belief state. If we view bargaining as mutual belief revision, then the agreement of bargaining, i.e., $F_1(G) + F_2(G)$, is exactly the common demands the agents accept each other, i.e., $(X_1 \otimes_1 F_2(G)) \cap (X_2 \otimes_2 F_1(G))$. In other words, the solution $F(G)$ should be the fixed-point with respect to the game $G$. The above theorem shows that this can be true if the demand sets of a game are logically closed.

To show this theorem, we need a few technical lemmas.

**Lemma 3** *For any bargaining game $G = ((X_1, \succeq_1), (X_2, \succeq_2))$,*

1. *$F_1(G) \subseteq X_1 \otimes_1 F_2(G)$;*

2. *$F_2(G) \subseteq X_2 \otimes_2 F_1(G)$.*

**Proof:** According to the definition of prioritized base revision, we have $X_1 \otimes_1 F_2(G) = \bigcap_{H \in X_1 \Downarrow F_2(G)} Cn(H \cup F_2(G))$. For any $H \in X_1 \Downarrow F_2(G)$, there is a deal $(D_1, D_2) \in \Omega(G)$ such that $D_1 = H$. This is because we can extend the pair $(H, F_2(G))$ to a deal $(H, D_2)$ such that $F_2(G) \subseteq D_2$. On the other hand, since $\Phi_1 \cup F_2(G)$ is consistent, we have $\Phi_1 \subseteq H$, where $(\Phi_1, \Phi_2)$ is the core of $G$. Thus, $\Phi_1 \subseteq D_1$ and $\Phi_2 \subseteq D_2$. It follows that $(D_1, D_2) \in \gamma(G)$. Since $F_1(G) \subseteq D_1$, we have $F_1(G) \subseteq H$. We conclude that $F_1(G) \subseteq X_1 \otimes_1 F_2(G)$. The proof of the second statement is similar. ¶

By this lemma we have,

1. $F_1(G) + F_2(G) \subseteq X_1 \otimes_1 F_2(G)$;

2. $F_1(G) + F_2(G) \subseteq X_2 \otimes_2 F_1(G)$.

Note that the above lemma does not require the demand sets $X_1$ and $X_2$ to be logically closed. However, without the assumption, the following lemmas do not hold.

**Lemma 4** *Let $(\Phi_1, \Phi_2)$ be the core of game $G = ((X_1, \succeq_1), (X_2, \succeq_2))$. If $X_1$ and $X_2$ are logically closed, then*

1. *$X_1 \otimes_1 F_2(G) = X_1 \otimes_1 (\Phi_2 + (X_1 \cap X_2))$;*

2. *$X_2 \otimes_2 F_1(G) = X_2 \otimes_2 (\Phi_1 + (X_1 \cap X_2))$*

**Proof:** We only present the proof of the first statement. The second one is similar. Firstly, we prove that $F_2(G) \subseteq \Phi_1 + \Phi_2 + (X_1 \cap X_2)$. If $X_1 \cup X_2$ is consistent, the result is obviously true. Therefore we can assume that $X_1 \cup X_2$ is inconsistent.

Assume that $\varphi \in F_2(G)$. If $\varphi \notin \Phi_1 + \Phi_2 + (X_1 \cap X_2)$, we have $\{\neg\varphi\} \cup \Phi_1 \cup \Phi_2 \cup (X_1 \cap X_2)$ is consistent. According to Lemma 2, we have $X_1^{\leq \pi_{max}^G + 1} \cup X_2^{\leq \pi_{max}^G + 1} \cup (X_1 \cap X_2)$ is inconsistent. Since our language is finite and both $X_1$ and $X_2$ are logically closed, the sets $X_1 \cap X_2$, $X_1^{\leq \pi_{max}^G + 1}$ and $X_2^{\leq \pi_{max}^G + 1}$ are all logically closed (the latter two due to LC). Therefore each set has a finite axiomatization. Let sentence $\psi_0$ axiomatize $X_1 \cap X_2$, $\psi_1$ axiomatize $X_1^{\leq \pi_{max}^G + 1}$ and $\psi_2$ axiomatize $X_2^{\leq \pi_{max}^G + 1}$. Thus $\psi_0 \wedge \psi_1 \wedge \psi_2$ is inconsistent. Notice that $\psi_0 \wedge \psi_1 \in X_1$ and $\psi_0 \wedge \psi_2 \in X_2$. It follows that $\neg\varphi \vee (\psi_0 \wedge \psi_1) \in X_1$ and $\neg\varphi \vee (\psi_0 \wedge \psi_2) \in X_2$. Since $\{\neg\varphi\} \cup \Phi_1 \cup \Phi_2 \cup (X_1 \cap X_2)$ is consistent, there is a deal $(D_1, D_2) \in \gamma(G)$ such that $\{\neg\varphi \vee (\psi_0 \wedge \psi_1)\} \cup \Phi_1 \cup (X_1 \cap X_2) \subseteq D_1$ and $\{\neg\varphi \vee (\psi_0 \wedge \psi_2)\} \cup \Phi_2 \cup (X_1 \cap X_2) \subseteq D_2$. We know that $\varphi \in F_2(G)$, so $\varphi \in D_1 + D_2$. Thus $\psi_0 \wedge \psi_1 \wedge \psi_2 \in D_1 + D_2$, which contradicts the fact that $D_1 + D_2$ is consistent. Therefore, we have shown that $F_2(G) \subseteq \Phi_1 + \Phi_2 + (X_1 \cap X_2)$.

Now we prove that $X_1 \otimes_1 F_2(G) = X_1 \otimes_1 (\Phi_2 + (X_1 \cap X_2))$. By Lemma 3, we have $\Phi_1 + \Phi_2 \subseteq X_1 \otimes_1 F_2(G)$. It follows that $X_1 \otimes_1 F_2(G) = (X_1 \otimes_1 F_2(G)) + (\Phi_1 + \Phi_2)$. Furthermore, we yield $X_1 \otimes_1 F_2(G) = (X_1 \otimes_1 F_2(G)) + (\Phi_1 + \Phi_2) + (X_1 \cap X_2)$ because $X_1 \cap X_2 \subseteq F_2(G)$. Since $F_2(G) \subseteq \Phi_1 + \Phi_2 + (X_1 \cap X_2)$. According to the AGM postulates, we have $(X_1 \otimes_1 F_2(G)) + (\Phi_1 + \Phi_2 + (X_1 \cap X_2)) = X_1 \otimes_1 (\Phi_1 + \Phi_2 + (X_1 \cap X_2))$. Therefore $X_1 \otimes_2 F_2(G) = X_1 \otimes_1 (\Phi_1 + \Phi_2 + (X_1 \cap X_2))$. In addition, it is easy to prove that $\Phi_1 \subseteq X_1 \otimes_1 (\Phi_2 + (X_1 \cap X_2))$. By the AGM postulates again, we have $X_1 \otimes_1 (\Phi_2 + (X_1 \cap X_2)) = (X_1 \otimes_1 (\Phi_2 + (X_1 \cap X_2))) + \Phi_1 = X_1 \otimes_1 (\Phi_1 + \Phi_2 + (X_1 \cap X_2))$. Therefore $X_1 \otimes_1 F_2(G) = X_1 \otimes_1 (\Phi_2 + (X_1 \cap X_2))$. ¶

The following lemma will complete the proof of Theorem 2.

**Lemma 5** *If $X_1$ and $X_2$ are logically closed, then*

$$(X_1 \otimes_1 F_2(G)) \cap (X_2 \otimes_2 F_1(G)) \subseteq F_1(G) + F_2(G).$$

**Proof:** Let $(\Phi_1, \Phi_2)$ be the core of $G$. Let

$$\Phi_1' = X_1^{\leq \pi_{max}^1} \text{ and } \Phi_2' = X_2^{\leq \pi_{max}^2}$$

where $\pi_{max}^1 = \max\{k : X_1^{\leq k} \cup \Phi_2 \cup (X_1 \cap X_2) \text{ is consistent}\}$ and $\pi_{max}^2 = \max\{k : \Phi_1 \cup X_2^{\leq k} \cup (X_1 \cap X_2) \text{ is consistent}\}$.

Note that in the cases when $\pi_{max}^i$ does not exist, we simply assume that it equals to $+\infty$. We claim that $X_1 \otimes_1 F_2(G) = \Phi_1' + F_2(G)$ and $X_2 \otimes_2 F_1(G) = \Phi_2' + F_1(G)$. We shall provide the proof of the first statement. The second one is similar.

Firstly, according to Lemma 2, $\Phi_1 \subseteq \Phi_1'$. Secondly, by Lemma 4, we have $X_1 \otimes_1 F_2(G) = X_1 \otimes_1 (\Phi_2 + (X_1 \cap X_2))$. Therefore to show $X_1 \otimes_1 F_2(G) = \Phi_1' + F_2(G)$, we only need to prove that $X_1 \otimes_1 (\Phi_2 + (X_1 \cap X_2)) = \Phi_1' + F_2(G) + (X_1 \cap X_2)$. This is because $\Phi_2 + (X_1 \cap X_2) \subseteq F_2(G)$, $F_2(G) \subseteq \Phi_1 + \Phi_2 + (X_1 \cap X_2)$ and $\Phi_1 \subseteq \Phi_1'$. By the construction of prioritized revision, we can easily verify that $\Phi_1' + \Phi_2 + (X_1 \cap X_2) \subseteq X_1 \otimes_1 (\Phi_2 + (X_1 \cap X_2))$. Therefore we only have to show the other direction, i.e., $X_1 \otimes_1 (\Phi_2 + (X_1 \cap X_2)) \subseteq \Phi_1' + \Phi_2 + (X_1 \cap X_2)$.

If $\Phi_1' = X_1$, then $X_1 \cup (\Phi_2 + (X_1 \cap X_2))$ is consistent. It follows that $X_1 \otimes_1 (\Phi_2 + (X_1 \cap X_2)) \subseteq X_1 + (\Phi_2 + (X_1 \cap X_2)) = \Phi_1' + \Phi_2 + (X_1 \cap X_2)$, as desired. If $\Phi_1' \neq X_1$, according to the definition of $\pi_{max}^1$, we have $X_1^{\leq \pi_{max}^1 + 1} \cup \Phi_2 \cup (X_1 \cap X_2)$ is inconsistent. Therefore there exists $\psi \in X_1^{\leq \pi_{max}^1 + 1}$ such that $\neg\psi \in \Phi_2 + (X_1 \cap X_2)$. Now we assume that

$\varphi \in X_1 \otimes_1 (\Phi_2 + (X_1 \cap X_2))$. If $\varphi \notin \Phi'_1 + \Phi_2 + (X_1 \cap X_2)$, then $\{\neg\varphi\} \cup \Phi'_1 \cup \Phi_2 \cup (X_1 \cap X_2)$ is consistent. So is $\{\neg\varphi \vee \psi\} \cup \Phi'_1 \cup \Phi_2 \cup (X_1 \cap X_2)$. Notice that $\neg\varphi \vee \psi \in X_1^{\leq \pi^1_{max}+1}$. There exists $H \in X_1 \Downarrow (\Phi_2 + (X_1 \cap X_2))$ such that $\{\neg\varphi \vee \psi\} \cup \Phi'_1 \subseteq H$. Since $\varphi \in X_1 \otimes_1 (\Phi_2 + (X_1 \cap X_2))$ and $H$ is logically closed, we have $\psi \in H$, which contradicts the consistency of $H \cup (\Phi_2 + (X_1 \cap X_2))$. Therefore $X_1 \otimes_1 (\Phi_2 + (X_1 \cap X_2)) \subseteq \Phi'_1 + \Phi_2 + (X_1 \cap X_2)$.

Finally we prove the claim of the lemma. Let $\varphi \in (X_1 \otimes_1 F_2(G)) \cap (X_2 \otimes_2 F_1(G))$. We then have $\varphi \in (\Phi'_1 + F_2(G)) \cap (\Phi'_1 + F_2(G))$. For $\varphi \in \Phi'_1 + F_2(G)$, there exists a sentence $\psi_2$ such that $F_2(G) \vdash \psi_2$ and $\varphi \vee \neg\psi_2 \in \Phi'_1$. Similarly, there exists a sentence $\psi_1$ such that $F_1(G) \vdash \psi_1$ and $\varphi \vee \neg\psi_1 \in \Phi'_2$. It turns out that $\varphi \vee \neg\psi_1 \vee \neg\psi_2 \in \Phi'_1 \cap \Phi'_2$. Thus $\varphi \vee \neg\psi_1 \vee \neg\psi_2 \in X_1 \cap X_2$. However, $X_1 \cap X_2 \subseteq F_1(G) + F_2(G)$. It follows that $\varphi \vee \neg\psi_1 \vee \neg\psi_2 \in F_1(G) + F_2(G)$. Note that $\psi_1 \wedge \psi_2 \in F_1(G) + F_2(G)$. Therefore we conclude that $\varphi \in F_1(G) + F_2(G)$. ¶

Theorem 2 establishes the link between bargaining theory and belief revision. The link helps us to understand the reasoning process behind bargaining. It is even more interesting if we can extend the result into the general multiagent case. However, the main challenge is how multiple agents mutually revise their beliefs.

## 5. Game-theoretic Properties of the Bargaining Solution

In game theory, the properties that are considered to be important to a bargaining solution include *individual rationality, Pareto optimality, ordinal invariance (or scale invariance), symmetry* and *contraction independence*. In our bargaining model, bargainers' preferences are represented in total preorder, any order-preserving transformation on the preferences does not change the order of preferences. Therefore our solution satisfies ordinal invariance trivially. In this section, we will examine the other properties in the above list with our bargaining solution. Before presenting the results, we first introduce a few concepts that are necessary for the game-theoretic analysis of bargaining.

### 5.1 Strategies and Utilities

Two concepts play essential roles in game-theoretic analysis of bargaining: *strategy* and *utility*. Given a bargaining game $G = ((X_1, \succeq_1), (X_2, \succeq_2))$, a *strategy profile* of the game is a pair $(S_1, S_2)$ where $S_1 \subseteq X_1$ and $S_2 \subseteq X_2$. The strategy profile can be interpreted as a pair of proposals of demands from both players in a course of bargaining.

We say a strategy profile $S = (S_1, S_2)$ to be *compatible* if

1. $X_1 \cap X_2 \subseteq S_1$ and $X_1 \cap X_2 \subseteq S_2$

2. $S_1 \cup S_2$ is consistent

Obviously any deal of a game is a compatible strategy profile. The bargaining solution $F(G)$ is also a compatible strategy profile of $G$.

Now we consider the gains of each player from a strategy profile. Assume that the strategy profile $(S_1, S_2)$ leads to an agreement, the player $i$'s payoff or utility is defined as:

$$u_i(S_i) = \begin{cases} \max\{k : X_i^{\leq k} \subseteq S_i\}, & \text{if } S_i \neq X_i; \\ \min\{k : X_i^{\leq k} = X_i\}, & \text{otherwise.} \end{cases}$$

In other words, $u_i(S)$ counts the number of top block demands that are covered by $S_i$. Note that the payoff does not count individual demands. Specifically we define the utility for the default disagreement point $(\emptyset, \emptyset)$ to be $(0, 0)$.

## 5.2 Pareto Optimality

Based on the above definition, it is easy to see that our solution satisfies individual rationality (in the sense of game theory) because for any game $G$, $u_i(F_i(G)) \geq 0$ for $i = 1, 2$. Now we consider Pareto efficiency.

Pareto optimality is one of the most important properties of bargaining solution. We call a compatible strategy profile $(S_1, S_2)$ of a game to be *(strictly) Pareto optimal* if there does not exist a compatible strategy profile $(S_1', S_2')$ of the game such that either $u_1(S_1') \geq u_1(S_1)$ & $u_2(S_2') > u_2(S_2)$ or $u_1(S_1') > u_1(S_1)$ & $u_2(S_2') \geq u_2(S_2)$.

A compatible strategy profile $(S_1, S_2)$ of a game is *weakly Pareto optimal* if there does not exist another compatible strategy profile $(S_1', S_2')$ of the game such that $u_1(S_1') > u_1(S_1)$ and $u_2(S_2') > u_2(S_2)$.

**Theorem 3** *For any bargaining game $G$, $F(G)$ is weakly Pareto optimal.*

**Proof:** Suppose that there is a compatible strategy profile $(S_1, S_2)$ of $G$ such that $u_1(S_1) > u_1(F_1(G))$ and $u_2(S_2) > u_2(F_2(G))$. Then $u_1(S_1) > \pi_{max}^G$ and $u_2(S_2) > \pi_{max}^G$. Since $(S_1, S_2)$ is compatible, $X_1^{\leq u_1(S_1)} \cup X_2^{\leq u_2(S_2)} \cup (X_1 \cap X_2)$ is consistent. It turns out that $\max\{k : X_1^k \cup X_2^k \cup (X_1 \cap X_2)$ is consistent$\} > \pi_{max}^G$, which contradicts Lemma 2. Therefore $F(G)$ is weakly Pareto optimal. ¶

Obviously, our solution $F$ does not satisfy strict Pareto optimality. For instance, the solution of $G'$ in Example 5 (the second part of the example) is not strictly Pareto optimal (but it is weakly Pareto optimal). This is not a problem of the solution but the nature of the problem domain we consider. It is well known in game theory that if the feasible set of a bargaining game is not convex, there is no guarantee of unique bargaining solution that is strictly Pareto optimal (Kaneko, 1980; Mariotti, 1996). Therefore, for non-convex domain, we need a trade-off between the uniqueness of solutions and strict Pareto optimality (Conley & Wilkie, 1991, 1996; Mariotti, 1998; Xu & Yoshihara, 2006). By using a similar approach introduced by Zhang (2007), we can map a logically represented bargaining game into a numerically represented bargaining game. Under such a mapping, the feasible set that corresponds to any logically represented bargaining game is non-convex unless the demand sets of the logical bargaining game is consistent. Since our solution is a unique solution, we cannot expect it to be strictly Pareto optimal.

## 5.3 Restricted Symmetry

In game theory, a bargaining game is symmetric if the feasible set is invariant under any permutation of each point in the feasible set (Nash, 1950). However, the concept of symmetry is not easy to be extended to the logic-based bargaining models because a bargaining problem is represented by its physical items. Permutation of deals does not make any sense. One may wonder how to judge the fairness of bargaining without the concept of symmetry. In our point of view, there is no such a thing as fair outcome in negotiation. The outcome of a negotiation relies on the bargaining power of each party. A bargainer with higher negotiation power receives more gains from the negotiation. However, it is reasonable to assume that any negotiation should be based on a fair bargaining procedure or a negotiation protocol. The construction of our bargaining solution is meant to capture the idea of fair negotiation protocols. The approach we use in this paper is similar to the idea of bargaining with an agenda (O'Neill et al., 2004). We consider a negotiation process consists of several rounds or stages. At each round, the parties are to reach agreements on new issues that have not been considered in the previous rounds. We assume that each party always place the higher wanted demands at earlier rounds. All the demands that have been mutually accepted in the earlier rounds have to remain in the agreements of the negotiation in the later rounds. Once there is no new agreement being reached, the negotiation procedure stops. With such a process, a negotiation always terminates at the same level of priority of demands for all players.

**Theorem 4** *For any bargaining game $G = ((X_1, \succeq_1), (X_2, \succeq_2))$, if $X_1$ and $X_2$ are logically closed, then there is a natural number $n$ such that*

$$\boldsymbol{F}(G) = (X_1 \cap (X_1^{\leq n} + X_2^{\leq n} + (X_1 \cap X_2)), X_2 \cap (X_1^{\leq n} + X_2^{\leq n} + (X_1 \cap X_2)))$$

**Proof:** In fact, we can prove that $n = \pi_{max}^G$. In such a case $X_1^{\leq n} + X_2^{\leq n} = \Phi_1 + \Phi_2$.

Obviously if $X_1 \cup X_2$ is consistent, then the result is trivial. Therefore we assume that $X_1 \cup X_2$ is inconsistent. We only prove the case that $F_1(G) = X_1 \cap (\Phi_1 + \Phi_2 + (X_1 \cap X_2))$. The proof of the other part is similar.

For any $(D_1, D_2) \in \gamma(G)$, we have $\Phi_1 \subseteq D_1$ and $\Phi_2 \subseteq D_2$. We prove that $X_1 \cap (\Phi_1 + \Phi_2 + (X_1 \cap X_2)) \subseteq D_1$. If it is not the case, there exists a sentence $\varphi \in X_1 \cap (\Phi_1 + \Phi_2 + (X_1 \cap X_2))$ such that $\varphi \notin D_1$. On the one hand, $\varphi \in X_1 \cap (\Phi_1 + \Phi_2 + (X_1 \cap X_2))$ implies that $D_1 \cup D_2 \vdash \varphi$ because $\Phi_1 + \Phi_2 + (X_1 \cap X_2) \subseteq D_1 + D_2$. On the other hand, $\varphi \notin D_1$ implies that $\{\varphi\} \cup D_1 \cup D_2$ is inconsistent due to the maximality of deals. It follows that $D_1 \cup D_2 \vdash \neg\varphi$. Therefore $D_1 \cup D_2$ is inconsistent, a contradiction. We have proved that for any deal $(D_1, D_2) \in \gamma(G)$, $X_1 \cap (\Phi_1 + \Phi_2 + (X_1 \cap X_2)) \subseteq D_1$. Thus $X_1 \cap (\Phi_1 + \Phi_2 + (X_1 \cap X_2)) \subseteq \bigcap_{(D_1, D_2) \in \gamma(G)} D_1 = F_1(G)$. The proof of Lemma 4 has shown that the other direction of inclusion $F_1(G) \subseteq X_1 \cap (\Phi_1 + \Phi_2 + (X_1 \cap X_2))$ holds. Therefore $F_1(G) = X_1 \cap (\Phi_1 + \Phi_2 + (X_1 \cap X_2))$. ¶

This theorem shows that at the termination of negotiation, each party can remain the demands down to the same level, i.e., $\pi_{max}^G$, including the common demands and their logical consequences. However, the solution seemingly excludes the low ranked irrelevant items. This is due to the assumption of the logical closedness on the demand sets. In fact, with this assumption, no items are irrelevant because for any two statements $\varphi, \psi \in X_i$, we always have $\varphi \to \psi$ and $\psi \to \varphi$ in $X_i$. This is why we do not assume logical closedness in general.

Another question that may arise is that a player could gain more negotiation power than the other if he puts more negotiation items in earlier stages of his agenda (effectively could gain more if the negotiation does not end up with disagreement). In fact, this is true and natural if the risk of breakdown is taken into account. If a player places a conflictive item in an earlier stage in his agenda, the negotiation would terminate sooner. Therefore the ordering of demands is a part of the strategy of a bargainer. We will discuss this issue in a separate section (see Section 6)

### 5.4 Contraction Independence

*Contraction Independence*, or called *Independence of Irrelevant Alternatives* (IIA), requires that if an alternative is judged to be the best compromise for some problem, then it should still be judged best for any subproblem that contains it (Thomson, 1994). For logic-based bargaining model, alternatives are not explicitly given. However, we can easily define the concept of subproblem in terms of bargainers' prioritized demand sets.

A bargaining game $G' = ((X_1', \succeq_1'), (X_2', \succeq_2'))$ is a subgame of $G = ((X_1, \succeq_1), (X_2, \succeq_2))$, denoted by $G' \sqsubseteq G$, if for any $i = 1, 2$,

1. $X_i' \subseteq X_i$,

2. $\succeq_i' = \succeq_i \cap (X_i' \times X_i')$,

3. for any $\varphi \in X_i$, if there is $\psi \in X_i'$ and $\varphi \succeq \psi$, then $\varphi \in X_i'$.

In other words, $X_i'$ is the upper segment of $X_i$ with respect to $X_i$'s hierarchy.

**Theorem 5** *Let $G' \sqsubseteq G$. If $F(G)$ is a strategy profile of $G'$, then $F(G') = F(G)$.*

**Proof:** First, since $F(G)$ is a strategy profile of $G'$, we have $X_1 \cap X_2 \subseteq X_i'$ ($i = 1, 2$). According to the definition of subgame, it is easy to show that $\pi_{max}^{G'} \leq \pi_{max}^{G}$. By the condition that $F(G)$ is a strategy profile of $G'$ again, we have $\pi_{max}^{G'} = \pi_{max}^{G}$, which means that $G'$ and $G$ share the same core $(\Phi_1, \Phi_2)$. For each deal $(D_1, D_2) \in \gamma(G)$, obviously $(D_1 \cap X_1', D_2 \cap X_2')$ is a deal of $G'$ and $(D_1 \cap X_1', D_2 \cap X_2') \in \gamma(G')$. It follows that $F_1(G') \subseteq D_1$ and $F_2(G') \subseteq D_2$. Therefore $F_1(G') \subseteq F_1(G)$ and $F_2(G') \subseteq F_2(G)$.

On the other hand, for each $(D_1', D_2') \in \gamma(G')$, we can extend it into a deal $(D_1, D_2)$ of $G$ such that $(D_1, D_2) \in \gamma(G)$ because $G'$ and $G$ share the same core and $X_1 \cap X_2 \subseteq X_i'$ ($i = 1, 2$). Since $F(G)$ is a strategy profile of $G'$, we have $F_1(G) \subseteq D_1 \cap X_1'$ and $F_2(G) \subseteq D_2 \cap X_2'$. It follows that $F_1(G) \subseteq D_1'$ and $F_2(G) \subseteq D_2'$, which implies $F_1(G) \subseteq F_1(G')$ and $F_2(G) \subseteq F_2(G')$. We conclude that $F(G') = F(G)$. ¶

Note that the claim of the above theorem is weaker than Nash's IIA because, for any two bargaining games $G$ and $G'$, the alternatives of $G'$ being a subset of the alternatives of $G$ do not guarantee $G'$ is a subgame of $G$. However, as it has been pointed out by many authors, Nash's IIA in his original form is no longer a plausible assumption for the domain of non-convex bargaining problems (Conley & Wilkie, 1996; Mariotti, 1998; Zhang, 2007).

## 6. Bargaining Power and Risk Posture

As we have shown in Section 1, representing bargainers' preferences in ordinal does not automatically solve the problem of ordinal bargaining (see Example 1). This is because ordinal preference has much less expressive power than cardinal utility. It is unable to express risk posture of a player, which determines the player's bargaining power. A successful solution to the ordinal bargaining problem should supply an alternative mean to express bargainers' attitude towards risk. In this section, we will illustrate with two case studies how this problem is solved using our framework.

## 6.1 Case Study I: Bargainer's Attitude towards Risk

Let us revisit the bargaining game in Example 4, where the demand sets are $X_1 = \{p, r\}$ and $X_2 = \{q, \neg r\}$. Consider the following variations of preferences, which reflect the difference of players' attitude towards risk (note that different cases lead to different games).

$G^1 : X_1 = \{\{p\}, \{r\}\}$ and $X_2 = \{\{q\}, \{\neg r\}\}$[10]

$G^2 : X_2 = \{\{p, r\}, \{\}\}$ and $X_2 = \{\{q\}, \{\neg r\}\}$

$G^3 : X_3 = \{\{p\}, \{r\}\}$ and $X_2 = \{\{q, \neg r\}, \{\}\}$

$G^4 : X_4 = \{\{p, r\}\}$ and $X_2 = \{\{q, \neg r\}\}$

It is easy to calculate the solutions of the games:

$F(G^1) = (\{p\}, \{q\})$

$F(G^2) = (\{p, r\}, \{q\})$

$F(G^3) = (\{p\}, \{q, \neg r\})$

$F(G^4) = (\emptyset, \emptyset)$

In $G^1$, both players are risk-averse, where the players rank the conflicting item $r$ being the lowest priority. This means that both have the incentive to reach an agreement. In $G^2$, player 1 is more aggressive since the conflicting item is highly ranked. We find that player 1 won the game in this case. This is not surprising. In general, a risk-averse player would gain disadvantage in negotiation comparing to a risk-lover (see Roth's book, 1979b, p.35-60). $G^3$ is symmetrical to $G^2$. In $G^4$, both players are aggressive, therefore the game ends up with disagreement.

From the example, it is clear to see how bargainers' attitude towards risk are specified in our model. A risk-averse player would give the demands that likely conflict with the demands of the other player relatively lower priorities so that an agreement is more likely to be reached. In contrast, a risk-loving player would more firmly entrench those conflicting demands. Notice that logic plays a crucial role in the representation. Simply expressing bargainers' demands in physical terms without specifying their relation is not sufficient to lead to an ordinal solution. It is crucial to specify the logical relations between the demands from all players. In the above example, the contradiction demands $r$ and $\neg r$ play the main role in the determination of the solutions. This is the main difference of our model from the game-theoretic models.

## 6.2 Case Study II: Discretization of Numerical Games

As we have mentioned in the introduction section, risk posture is represented in non-linearity of utility functions in Nash's model. In this subsection, we use the example of cake division (Example 1) to show how the non-linearity of preferences is represented in our model.

---

10. The demand sets are represented in prioritized partitions. Their corresponding preference orderings are $p \succeq_1 r$ and $q \succeq_2 \neg r$, respectively.

To represent the bargaining problem in logical form, we need to discretize the domain. Let $p_n = \{$*Player A receives no less than n percentages of the cake and player B gets the remain*$\}$, where $n$ is a natural number between 0 to 100. In addition, the following constraints should be acknowledged by both players:

$$C = \{p_{n+1} \to p_n : n = 0, 1, 2, \cdots, 100\}$$

which says that if player 1 receives no less than $n + 1$ percents of the cake, then he must receive no less than $n$ percents.

Then the demands of two players can be represented as

$X_A = C \cup \{p_0, p_1, p_1, \cdots, p_{100}\}$

$X_B = C \cup \{\neg p_{101}, \neg p_{100}, \neg p_{99}, \cdots, \neg p_1\}$

Assume that player A arranges his demands according to the linear scale of his share. The hierarch of his demand set is[11]:

$$\{C, \{p_0 \cdots, p_5\}, \{p_6 \cdots, p_{10}\}, \cdots, \{p_{96}, \cdots, p_{100}\}\}$$

Player B arranges her demands according to the square of her share. The hierarch of her demand set is[12]:

$$\{C, \{\neg p_{101}, \cdots, \neg p_{79}\}, \{\neg p_{78}, \cdots, \neg p_{70}\}, \cdots, \{\neg p_6, \neg p_5, \neg p_4\}, \{\neg p_3, \neg p_2\}, \{\neg p_1\}\}$$

According to the above setting, the solution of the bargaining game is:

$$(C \cup \{p_{34}, p_{33}, \cdots, p_0\}, C \cup \{\neg p_{42}, \neg p_{43}, \cdots, \neg p_{101}\})$$

It is easy to calculate that the players agree on the division of the cake to be $34 \le x_A < 42$ and $58 < x_B \le 66$.[13] Therefore our solution gives similar prediction as the game-theoretic solutions. This indicates that the risk posture of the players has been embedded in our model. In fact, we can easily see that player B is more aggressive because he ranks higher conflicting items relatively higher than player A does. For instance, player B ranks the equal share (50/50) at the 7th level while player A ranks it at 11th level.

As we have seen again from the example, the ordering of demands does not reflect player's gains from the demands but represents the player's preference of retaining or abandoning his demands. This is another significant difference between our bargaining model and the game-theoretic models.

One may ask that whether a player could get advantages by "cheating" in the sense that if an agent knows the demands and ranking of the other party, the agent can adjust her demand hierarchy in order to obtain a better outcome. Yes, it is possible. You can tell your opposite what you want (your demands) but you should not release your ranking on your demands. Otherwise, you lose your bargaining power. The reason is that your attitude towards risk has been encoded in your ranking on your demands.

---

11. This indicates that player A ignores small differences of divisions. For instance, he may consider that any share between 0-5% means the same to him. In the real negotiation, the player may request 100%, 95%, $\cdots$ 5% in sequence by giving up 5% at each round in a bargaining.

12. Player B claims his share in the sequence 100%, 98%, 95%, $\cdots$ by dropping his demand in the scale of square.

13. Note that there is no communication between the players. Therefore a player may give up more then it is needed (similar to sealed-bid auction).

## 7. Computational Complexity

In this section, we study the computational properties of the bargaining solution we developed earlier. We assume that readers are familiar with the complexity classes of P, NP, coNP, $\Sigma_2^P$ and $\Pi_2^P = co\Sigma_2^P$. The class of DP contains all languages $L$ such that $L = L_1 \cap L_2$ where $L_1$ is in NP and $L_2$ is in coNP. Also the class $\Delta_{k+1}^P = P^{\Sigma_k^P}$ contains all languages recognizable in polynomial time by a deterministic Turing machine with a $\Sigma_k^P$ oracle. In particular, the class $\Delta_2^P[\mathcal{O}(logn)]$ contains all languages recognizable by a deterministic Turing machine with $\mathcal{O}(logn)$ calls to an NP oracle[14] and $\Delta_3^P$ contains all languages recognisable by a deterministic Turing machine with a $\Sigma_2^P$ oracle. It is well known that $P \subseteq NP \subseteq DP \subseteq \Delta_2^P \subseteq \Sigma_2^P \subseteq \Delta_3^P$, and these inclusions are generally believed to be proper (readers may refer to the work of Papadimitriou, 1994, for further details).

Consider a bargaining game $G = ((X_1, \succeq_1), (X_2, \succeq_2))$ where $X_1$ and $X_2$ are finite. As we have mentioned in Section 2.3, we can always write $X_i = X_i^1 \cup \cdots \cup X_i^m$, where $X_i^k \cap X_i^l = \emptyset$ for any $k \neq l$. Also for each $k < m$, if a formula $\varphi \in X_i^k$, then there does not exist a $\psi \in X_i^l$ $(k < l)$ such that $\varphi \prec_i \psi$. Therefore, for the convenience of analysis, in the rest of this section, we will specify a bargaining game as $G = (X_1, X_2)$, where $X_1 = \bigcup_{i=1}^{m} X_1^i$ and $X_2 = \bigcup_{j=1}^{n} X_2^j$, and $X_1^1, \cdots, X_1^m$, and $X_2^1, \cdots, X_2^n$ are the partitions of $X_1$ and $X_2$ respectively and satisfy the properties mentioned above.

The following four major decision problems are most important in order to understand the computational properties for our bargaining game model developed in the previous sections: let $G = (X_1, X_2)$ be a bargaining game, we would like to decide: (1) whether a given pair $(D_1, D_1)$ where $D_i \subseteq X_i$ $(i = 1, 2)$ is a deal of $G$; (2) whether a given pair of propositional formulas $(\Phi_1, \Phi_2)$ is a core of $G$ respectively; (3) whether a given formula is derivable from the core of $G$; and (4) whether a given strategy profile of $G$ is a solution of $G$, First, we have the following result for deciding a deal for a given bargaining game.

**Theorem 6** *Let $G = (X_1, X_2)$ be a bargaining game, and $D_1 \subseteq X_1$ and $D_2 \subseteq X_2$. Deciding whether $(D_1, D_2)$ is a deal of $G$ is DP-complete.*

**Proof:** *Membership proof.* According to Definition 2, to decide whether $(D_1, D_2)$ is a deal of $G$, for $D_1$ (or $D_2$), we need to check: (1) for each $k = 1, \cdots, m$ (or for $k' = 1, \cdots, n$ resp.), whether $D_2 \cup \bigcup_{j=1}^{k}(D_1 \cap X_1^j)$ (or $D_1 \cup \bigcup_{j=1}^{k'}(D_2 \cap X_2^j)$ resp.) is consistent; (2) checking whether $X_1 \cap X_2 \subseteq D_i$ $(i = 1, 2)$; and (3) such $D_1$ and $D_2$ are maximal such subsets of $X_1$ and $X_2$ respectively. For (1), we observe that for each $k$, the set $\bigcup_{j=1}^{k}(D_1 \cap X_1^j)$ can be computed in polynomial time, and checking the consistency of $D_2 \cup \bigcup_{j=1}^{k}(D_1 \cap X_1^j)$ is in NP. The same for $D_2$ case. It is obvious to see that (2) can be done in polynomial time. Now we consider (3). In order to check whether $D_1$ and $D_2$ are the maximal subsets of $X_1$ and $X_2$ respectively satisfying the condition, we consider the complement the problem: assume that $D_1$ (we can also assume $D_2$) is not the maximal subset of $X_1$ satisfying the required conditions, then there exists some $k$ and some $\varphi \in (X_1 \setminus D_1)$ such that $D_2 \cup \bigcup_{j=1}^{k} ((D_1 \cup \{\varphi\}) \cap X_1^k)$ is consistent. Clearly, we can guess such $k$, formula $\varphi$ and an interpretation $S$, then check whether $S$ is a model of $D_2 \cup \bigcup_{j=1}^{k} ((D_1 \cup \{\varphi\}) \cap X_1^k)$. Obviously, this is in NP. So the original problem is in coNP.

---

14. Note that in literatures, different notions have been used to denote this complexity class such as $P^{NP[logn]}$, $P_{||}^{NP}$ and $\Theta_2^P$.

*Hardness proof.* It is known that for given propositional formulas $\varphi_1$ and $\varphi_2$, deciding whether $\varphi_1$ is satisfiable and $\varphi_2$ is unsatisfiable is DP-complete (Papadimitriou, 1994). Given two propositional formulas $\varphi_1$ and $\varphi_2$, we construct in polynomial time a transformation from the $\varphi_1$'s satisfiability and $\varphi_2$'s unsatisfiability to a deal decision problem of a game. We simply define a game $G = (X_1, X_2) = (\{\neg\varphi_2 \to \varphi_1 \wedge p\}, \{q\} \cup \{\neg p\})$, where $p, q$ are propositional atoms not occurring in $\varphi_1$ and $\varphi_2$. Note that $X_1 \cap X_2 = \emptyset$. Let $D_1 = X_1$ and $D_2 = \{q\}$. Now we show that $(D_1, D_2)$ is a deal of $G$, that is, $D_2$ are the maximal subset of $X_2$ such that $X_1 \cup D_2$ is consistent, if and only if $\varphi_1$ is satisfiable and $\varphi_2$ is unsatisfiable.

($\Rightarrow$) Clearly, if $\varphi_1$ is satisfiable and $\varphi_2$ is unsatisfiable, then $X_1 \cup D_2$ is consistent, but $X_1 \cup X_2$ is not consistent. So $(D_1, D_2)$ is a deal of $G$.

($\Leftarrow$) Suppose $\varphi_1$ and $\varphi_2$ are unsatisfiable. Then $X_1$ itself is not consistent. If both $\varphi_1$ and $\varphi_2$ are satisfiable, then it is observed that $X_1 \cup X_2$ is consistent. So $(D_1, D_2)$ is not a deal of $G$. Finally, suppose $\varphi_1$ is unsatisfiable and $\varphi_2$ is satisfiable. In this case, $X_1 \cup X_2$ is still consistent. That means, $(D_1, D_2)$ is not a deal of $G$ either. ¶

As we can see from the definition, the core of a bargaining plays an essential role in the construction of the bargaining solution. The following theorem provides the complexity result of its decision problem.

**Theorem 7** *Let $G$ a bargaining game. Deciding whether a given pair of sets of propositional formulas $(\Phi_1, \Phi_2)$ is the core of $G$ is DP-complete.*

**Proof:** *Membership proof.* Let $G = (X_1, X_2)$, where $X_1 = \bigcup X_1^i$, and $X_1 = \bigcup X_1^j$. We outline an algorithm to check whether $(\Phi_1, \Phi_2)$ is the core of $G$: (1) check whether for some $k$, $\Phi_1 = \bigcup_{i=1}^{k} X_1^i$ and $\Phi_2 = \bigcup_{i=1}^{k} X_2^i$, and $\Phi_1 \neq \bigcup_{i=1}^{k+1} X_1^i$ or $\Phi_2 \neq \bigcup_{i=1}^{k+1} X_2^i$; (2) check whether $\Phi_1 \cup \Phi_2$ is consistent; and (3) check if $\Phi_1 \cup X_1^{k+1} \cup \Phi_2 \cup X_2^{k+1}$ is not consistent. Clearly, (1) can be done in polynomial time, checking (2) is in NP and checking (3) is in coNP. So the problem is in DP.

*Hardness proof.* The hardness proof is similar to that as described in the proof of Theorem 7 with some variations. Given two propositional formulas $\varphi_1$ and $\varphi_2$, we reduce the decision problem of $\varphi_1$'s satisfiability and $\varphi_2$'s unsatisfiability to our problem. Let $G = (X_1, X_2)$, where $X_1 = \{\neg\varphi_2 \to \varphi_1 \wedge p\} \cup \{q\}$, and $X_2 = \{q\} \cup \{\neg p\}$, where $p, q$ are propositional atoms not occurring in $\varphi_1$ and $\varphi_2$. We specify a pair of sets of formulas: $(\Phi_1, \Phi_2) = (\{\neg\varphi_2 \to \varphi_1 \wedge p\}, \{q\})$.

Now we show that $(\Phi_1, \Phi_2)$ is the core of $G$ if and only if $\varphi_1$ is satisfiable and $\varphi_2$ is unsatisfiable. Suppose that $\varphi_1$ is satisfiable and $\varphi_2$ is unsatifiable. Then $\Phi_1 \cup \Phi = \{\varphi_1 \wedge p, q\}$, which is consistent. On the other hand, $\Phi_1 \cup \{q\} \cup \Phi_2 \cup \{\neg p\} = \{\varphi_1 \wedge p, q, \neg p\}$, which is not consistent. So $(\Phi_1, \Phi_2)$ is the core of $G$.

We prove the other direction. (1) Both $\varphi_1$ and $\varphi_2$ are satisfiable. Then $\Phi_1 \cup \Phi_2 = \{(\varphi_2 \vee \varphi_1) \wedge (\varphi_2 \vee p), q\}$, which is consistent. However, we can also see that $\Phi_1 \cup \{q\} \cup \Phi_2 \cup \{\neg p\}$ has at leat one model which satisfies $\varphi_2$, $q$ and $\neg p$. This implies that $(\Phi_1, \Phi_2)$ is no longer the core of $G$. (2) Both $\varphi_1$ and $\varphi_2$ are unsatisfiable. In this case, $\Phi_1 \cup \Phi_2$ is not consistent any more. So $(\Phi_1, \Phi_2)$ is not the core of $G$. (3) $\varphi_1$ is unsatisfiable and $\varphi_2$ is satisfiable. Again, under this situation, $\Phi_1 \cup \Phi_2$ is no longer consistent, and hence it is not the core of $G$. This completes our proof. ¶

Recall that the intuition behind the core is that the final agreement maximizes fairly each agent's demands without violating the overall consistency. Then it is interesting to know whether certain information is derivable from the agent's demands that are in the final agreement. Let $\Phi = (\Phi_1, \Phi_2)$ is the core of bargaining game $G$. We define $C(G) \vdash \varphi$ if and only if $\Phi_1 \vdash \varphi$ or $\Phi_2 \vdash \varphi$.

**Theorem 8** *Given a bargaining game $G$ and a propositional formula $\varphi$. Deciding whether $C(G) \vdash \varphi$ is $\Delta_2^P[\mathcal{O}(log n)]$-complete.*

**Proof:** Membership proof. We outline an algorithm of deciding $C(G) \vdash \varphi$ as follows: (1) compute the core $(\Phi_1, \Phi_2)$ of $G$, and (2) checking if $\Phi_1 \vdash \varphi$ or $\Phi_2 \vdash \varphi$. From the definition, $\Phi = (\Phi_1, \Phi_2)$ is the core of $G$ iff $\Phi_1 = \bigcup_{j=1}^{k} X_1^j$, and $\Phi_2 = \bigcup_{j=2}^{k} X_2^j$, where $k$ is the maximal number that makes $\Phi_1 \cup \Phi_2$ consistent. Clearly, such $k$ can be determinated by binay search with $\mathcal{O}(log n)$ NP oracle calls. Then checking $\Phi_1 \vdash \varphi$ or $\Phi_2 \vdash \varphi$ can be done with two NP oracle calls. So the problem is in $\Delta_2^P[\mathcal{O}(log n)]$.

Hardness proof. We reduce the $\Delta_2^P[\mathcal{O}(log n)]$-complete $PARITY_\omega^B$ (Kobler, Schoning, & Wagner, 1987; Wagner, 1988) to our problem. An instance of $PARITY_\omega^B$ is a set of propositional formulas $\varphi_1, \cdots, \varphi_n$ such that if $\varphi_i$ is not satisfiable, then for each $j \geq i$, $\varphi_j$ is not satisfiable. The problem is to decide whether the number of satisfiable formulas is odd. Without loss of generality, we assume $n$ is an even number. Then we construct in polynomial time a bargaining game $G = (X_1, X_2)$ as follows:

$$X_1 = \bigcup X_1^{n/2} = \{\neg \varphi_2 \to p\} \cup \{\neg \varphi_4 \to \varphi_3 \wedge p\} \cup \cdots \{\neg \varphi_n \to \varphi_{n-1} \wedge p\},$$
$$X_2 = \bigcup X_2^{n/2} = \{q_1\} \cup \{q_2\} \cup \cdots \cup \{q_{n/2}\},$$

where $p, q_1, \cdots, q_{n/2}$ are propositional atoms not occurring in $\varphi_1, \cdots, \varphi_n$. Let $\varphi = p$. Now we show that $C(G) \vdash p$ if and only if there is an odd number of satisfiable formulas in $\varphi_1, \cdots, \varphi_n$. First, suppose $k$ is an odd number, $\varphi_1, \cdots, \varphi_k$ are satisfiable, and $\varphi_{k+1}, \cdots, \varphi_n$ are not satisfiable. Then it is observed that $\Phi = \bigcup_{j=1}^{[k/2]} X_1^j \cup X_2^j = \{\neg \varphi_2 \to p, \neg \varphi_4 \to \varphi_3 \wedge p, \cdots, \neg \varphi_{k+1} \to \varphi_k \wedge p\} \cup \{q_1, \cdots, q_{[k/2]}\}$ is consistent, and $\Phi \cup \{\neg \varphi_{k+3} \to \varphi_{k+2} \wedge p\} \cup \{q_{[k/2]+1}\}$ is not consistent. So $(\bigcup_{j=1}^{[k/2]} X_1^j, \bigcup_{j=1}^{[k/2]} X_2^j)$ is the core of $G$. Also, since $\neg \varphi_{k+1}$ is not satisfiable, it follows that $\neg \varphi_{k+1} \to \varphi_k \wedge p$ is reduced to $\varphi_k \wedge p$, which is contained in $\bigcup_{j=1}^{[k/2]} X_1^j$. So $C(G) \vdash p$. Second, we assume that there is an even number of satisfiable formulas in $\varphi_1, \cdots, \varphi_n$. Let $k$ be the even number such that $\varphi_1, \cdots, \varphi_k$ are satisfiable and $\varphi_{k+1}, \cdots, \varphi_n$ are not satisfiable. In this case, it can be observed that $\Phi = \bigcup\limits_{j=1}^{k/2} X_1^j \cup X_2^j = \{\neg\varphi_2 \rightarrow p, \neg\varphi_4 \rightarrow \varphi_3 \wedge p, \cdots, \neg\varphi_k \rightarrow \varphi_{k-1} \wedge p\} \cup \{q_1, \cdots, q_{k/2}\}$ is consistent, while $\Phi \cup \{\neg\varphi_{k+2} \rightarrow \varphi_{k+1} \wedge p\} \cup \{q_{k/2+1}\}$ is not consistent. So $(\bigcup\limits_{j=1}^{k/2} X_1^j, \bigcup\limits_{j=1}^{k/2} X_2^j)$ is the core of $G$. Then it is obvious that $p$ cannot be derived from $\bigcup\limits_{j=1}^{k/2} X_1^j$. That is, $C(G) \not\vdash p$. This completes our proof. ¶

Finally, we consider the decision problem for the solution of a given bargaining game. The following theorem gives its complexity upper bound.

**Theorem 9** *Let $G$ be a bargaining game, and $(S_1, S_2)$ a strategy profile of $G$. Deciding whether $(S_1, S_2)$ is the solution of $G$ is in $\Delta_3^P$.*

**Proof:** From Definition 3, we need to check whether $S_1 = \bigcap\limits_{(D_1, D_2) \in \gamma(G)} D_1$ and $S_2 = \bigcap\limits_{(D_1, D_2) \in \gamma(G)} D_2$, where $\gamma(G) = \{(D_1, D_2) \in \Omega(G) : \Phi_1 \subseteq D_1, \Phi_2 \subseteq D_2\}$, and $(\Phi_1, \Phi_2)$ is the core of $G$. Note that for simplicity, here we use the notion of core to represent the solution.

We first consider the complement of deciding $S_1 = \bigcap\limits_{(D_1, D_2) \in \gamma(G)} D_1$: checking whether $S_1 \neq \bigcap\limits_{(D_1, D_2) \in \gamma(G)} D_1$. Clearly, $S_1 \neq \bigcap\limits_{(D_1, D_2) \in \gamma(G)} D_1$ iff (1) $S_1 \nsubseteq \bigcap\limits_{(D_1, D_2) \in \gamma(G)} D_1$; or (2) $\bigcap\limits_{(D_1, D_2) \in \gamma(G)} D_1 \nsubseteq S_1$. We first guess a pair of sets of propositional formulas $(\Phi_1, \Phi_2)$, and check if it is the core of $G$. According to Theorem 8, we know that this is in $\Sigma_2^P$. Clearly, (1) holds iff there exists a formula $\varphi$ such that (a) $\varphi \in S_1$, and (b) $\varphi \notin \bigcap\limits_{(D_1, D_2) \in \gamma(G)} D_1$. Further, (b) holds iff there exists a deal $(D_1, D_2)$ such that $\Phi_1 \subseteq D_1, \Phi_2 \subseteq D_2$ and $\varphi \notin D_1$. Now we guess $\varphi$ and $(D_1, D_2)$, and then check if $\varphi \in S_1$, $(D_1, D_2)$ is a deal containing the core, and $\varphi \notin D_1$. From Thorem 7 we know that checking $(D_1, D_2)$ is a deal can be done with two NP oracle calls. So task (1) can be solved in $\Sigma_2^P$.

On the other hand, (2) holds iff there exists some $\varphi$ such that $\varphi \in \bigcap\limits_{(D_1, D_2) \in \gamma(G)} D_1$ and $\varphi \notin S_1$. To solve task (2), we consider its complement: for all $\varphi$, if $\varphi \notin S_1$, then $\varphi \notin \bigcap\limits_{(D_1, D_2) \in \gamma(G)} D_1$. Since there are only $|X_1| + |X_2|$ formulas we need to check, checking all $\varphi$ that are not in $S_1$ can be done in linear time. Then for each $\varphi \notin S_1$, we need to check if $\varphi \notin \bigcap\limits_{(D_1, D_2) \in \gamma(G)} D_1$, which, as shown above, can be done in $\Sigma_2^P$. Therefore, for all $\varphi$ that are not in $S_1$, there are at most $|X_1| + |X_2|$ checkings of whether $\varphi \notin \bigcap\limits_{(D_1, D_2) \in \gamma(G)} D_1$, which is in $P^{\Sigma_2^P} = \Delta_3^P$. This follows that deciding whether $S_1 = \bigcap\limits_{(D_1, D_2) \in \gamma(G)} D_1$ is in $\Delta_3^P$. Consequently, the original problem is also in $\Delta_3^P$. ¶

From the proof of Theorem 10, it can be observed that the computation for a bargaining game is very different from that of Nebel's prioritized belief revision.

**Theorem 10** *Let $G$ be a bargaining game and $(S_1, S_2)$ a strategy profile of $G$. Deciding whether $(S_1, S_2)$ is the solution of $G$ is DP-hard.*

**Proof:** We consider a special game $G = (X_1, X_2)$ where $X_1 = X_1^1 \cup X_1^2$, and $X_2 = X_2^1$ (i.e. $X_2$ has a singleton partition), and $X_1^1 \cup X_2^1$ is consistent but $X_1 \cup X_2$ is not consistent. In this case, we know that $(X_1^1, X_2^1)$ is the core of $G$. So our question is reduced to decide whether $S_2 = X_2$, and $S_1$ is the maximal subset of $X_1$ containing $X_1^1$ such that $S_1 \cup X_2$ is consistent. From the proof of Theorem 3, it is easy to see that this is DP-hard. ¶

Obviously, there is a gap between the lower bound and upper bound for the solution decision problem. This also sheds a light that our bargaining solution cannot be represented in terms of the traditional belief revision operators.

**Theorem 11** *Let $G$ be a bargaining game and $(S_1, S_2)$ a strategy profile of $G$. Deciding whether $(S_1, S_2)$ is the solution of $G$ is in $\Delta_2^P[\mathcal{O}(logn)]$, given that the set of all deals $\Omega(G)$ is provided.*

**Proof:** To decide whether $(S_1, S_2)$ is a solution of $G$, we need to do the following: (1) compute the core $(\Phi_1, \Phi_2)$ of $G$; (2) compute $\gamma(G)$ from $\Omega(G)$ and $(\Phi_1, \Phi_2)$; (3) compute $D_1' = \bigcap_{(D_1, D_2) \in \gamma(G)} D_1$ and $D_2' = \bigcap_{(D_1, D_2) \in \gamma(G)} D_2$; and (4) compare whether $S_1 = D_1'$ and $S_2 = D_2'$. Task (1) can be solved with one $\Delta_2^P[\mathcal{O}(logn)]$ oracle call; having $(\Phi_1, \Phi_2)$, task (2) can be done in polynomial time; and tasks (3) and (4) can be done in polynomial time. So the problem is in $\Delta_2^P[\mathcal{O}(logn)]$. ¶

## 8. Related Work

The investigation of ordinal bargaining theory diverges in two directions. The first direction focuses on the existence of ordinal solution in the Nash bargaining model. As we have mentioned in the introduction section, Shapley, Kibris, Safra and Samet have shown that there is a solution to the bargaining problems with three agents or more, which satisfies ordinal invariance, symmetry and Pareto optimality (Shubik, 1982; Özgür Kibris, 2004; Safra & Samet, 2004). This result is interesting because it shows a difference between bilateral bargaining and multilateral bargaining. However, it does not solve the problem of ordinal bargaining because the solution is not constructive and, more importantly, no alternative mean is offered to facilitate the representation of bargainers' risk attitude. Calvo and Perers investigated the bargaining problems with mixed players: cardinal players and ordinal players (Calvo & Peters, 2005). A bargaining solution is called *utility invariant* if it is ordinally invariant for the ordinal players and cardinally invariant for the cardinal players. It is proved that there is a solution satisfying utility invariance, Pareto optimality and individual rationality provided at least one player is cardinal. Obviously, this result is only peripheral because a utility invariant solution is not necessarily ordinal invariant.

The other direction of the investigation tries to circumvent Shapley's impossibility result by altering the Nash bargaining model. Rubinstein et al. reinterpreted the Nash bargaining solution with respect to ordinal preferences (Rubinstein et al., 1992). By restating Nash's axioms, it is proved that the redefined solution, referred to as ordinal-Nash solution, satisfies Pareto optimality, symmetry and contraction independence (ordinal invariance holds trivially). However, the result is based on the assumption that the preference ordering of each player is complete, transitive and continuous on the set of finite lotteries over a topological space. It is unclear that how to use such a specific preference language to describe players' risk attitudes. More important, the advantages of ordinal preferences, such as ease of elicitation and robustness of corresponded solutions, may be lost once the preference ordering is extended to the space of lotteries. O'Neill et al. model a bargaining situation with a family of Nash bargaining games, parameterized by time (O'Neill et al., 2004). A bargaining solution is then defined as a function that specifies an outcome at each time. With the model, bargainers' risk attitude can be expressed through varying preferences over time, which is very intuitive. Since a solution is no longer a single point but a function over time, the construction of the proposed ordinal solution relies on the solution of sets of simultaneous differential equations.

This work is developed based on a series of previous work of the authors. The fixed-point condition discussed in this paper was firstly proposed by Zhang et al. (2004)[15]. Zhang and Zhang (2006a, 2006b) presented a logical solution to the bilateral bargaining problem based on ordinal preferences. However, that solution does not satisfy the fixed point condition. Zhang (2008) showed that a revision of the solution satisfies the fixed-point condition. The present paper further develops and systemizes the solution and discusses its logical and game-theoretic properties.

## 9. Conclusion

We have presented a bargaining solution to the bilateral bargaining problem based on the logical representation of bargaining demands and ordinal representation of bargainers' preferences. We have shown that the solution satisfies most desirable logical properties, such as *individual rationality* (logical version), *consistency* and *collective rationality*, and the desirable game-theoretic properties, such as *weak Pareto optimality*, *restricted symmetry* and *contraction invariance*. The ordinal invariance and game-theoretic version of individual rationality hold trivially. Due to the discrete nature of logical representation, the solution is not (strictly) Pareto optimal and does not satisfy symmetry. However, if the demand sets of two players are logically closed, the solution meets a restricted version of symmetry. In addition, we have demonstrated that under the logical closedness assumption, the outcome of a negotiation is the result of mutual belief revision in terms of Nebel's prioritized belief base revision. This result established a link between bargaining theory and belief revision. Such a link would play an important role in the future on the research of multiagent belief revision and logic-based bargaining theory. Our complexity analysis indicates that the computation of bargaining solution is more difficult than prioritized belief base revision.

A satisfactory model of bargaining should be able to encode the key factors that determine the bargaining outcome, such as bargainer's demands, preferences, attitudes towards risk and so on. Cardinal utility specifies two sorts of information: *preference over possible agreements* (via the ordering of utility values) and *risk attitudes* (via the non-linearity of utility function). However, the second sort of information, which determines players' bargaining power, may be lost after an ordinal transformation. Meanwhile, a bargaining model based purely on ordinal information about preferences does not automatically solve the problem because bargainers' risk attitude is even inexpressible in such a model. Therefore an ordinal bargaining theory must supply a facility to describe the information other than ordinal preferences, including risk attitudes. In this paper, we specify a bargaining situation in logical structure. Bargainer's demands, goals and beliefs are described in logical statements. The conflicts of demands between two players can then be identified through consistency checking. More importantly, bargainer's attitudes towards risk are expressible in our model in a natural way: *a risk-averse player tends to give a conflicting demand a relatively lower priority so that an agreement could be more likely reached while a risk-lover would firmly entrench her demands with less care about whether her demands contradict her opponent's.*

---

15. Jin et al. (2007) also introduced a fixed-point condition for negotiation functions, which says that under certain conditions, negotiating on the outcome of a negotiation generates the same outcome. Obviously our bargaining solution satisfies this fixed-point condition because any outcome of bargaining is consistent, which remains the same in any further negotiation.

A few issues are worth further investigation. Firstly, we have shown that our solution satisfies a set of logical properties and game-theoretic properties. It is interesting to know whether there is an axiomatic system that exactly characterizes the solution. The main challenge here is that the construction of our solution is syntax-dependent. If we simply impose the logical closedness on the demand sets, we will lose a few desirable properties, such as the inclusion of irrelevant items and computational results. If we do not apply the assumption, we shall need the axioms to specify the way of logical representation. In other words, the axioms have to specify how a demand set should be represented syntactically.

Secondly, the present work offers a solution to the bilateral bargaining situations. It does not supply a model to bargaining agents. Therefore the current framework does not deal with the issues like "how a demand is formed?", "why a demand should be ranked higher than another?" or "how to bargain effectively?". It is interesting how a logic of agency can be used or developed to model bargaining agents and how such a logic interacts with the logic of bargaining.

Finally, a few issues on the computational complexity of the proposed solution remain unsolved. As we have shown in this paper the membership checking of the solution is DP-hard but in $\Delta_3^P$. It is not clear yet how this upper bound and lower bound gap can be closed. We think some new complexity proof technique may be needed for this challenge.

## Acknowledgments

The authors wish to thank Norman Foo, Michael Thielscher and the anonymous reviewers for their comments. This work was partially supported by the Australian Research Council with project LP0883646.